\documentclass[amsmath,amssymb]{revtex4}
\usepackage{graphicx}
\usepackage{amscd,amsmath,amsthm,amsfonts,amssymb}
\def\PT{$\cal{PT}$}
\def\[{\begin{equation}}
\def\]{\end{equation}}

\begin{document}
\title{On general rogue waves in the parity-time-symmetric nonlinear Schr\"{o}dinger equation}
\author{Bo Yang and Jianke Yang}
\affiliation{Department of Mathematics and Statistics, University of Vermont, Burlington, VT 05405, USA}

\begin{abstract}
This article addresses the question of general rogue-wave solutions in the nonlocal parity-time-symmetric nonlinear Schr\"{o}dinger equation. By generalizing the previous bilinear Kadomtsev-Petviashvili reduction method, large classes of rogue waves are derived as Gram determinants with Schur polynomial elements. It is shown that these rogue waves contain previously reported ones as special cases. More importantly, they contain many new rogue wave families. It is conjectured that the rogue waves derived in this article are \emph{all} rogue-wave solutions in the parity-time-symmetric nonlinear Schr\"{o}dinger equation.
\end{abstract}

\maketitle

\section{Introduction}
Rogue waves are rational solutions of nonlinear wave equations that ``appear from nowhere and
disappear with no trace" \cite{Akhmediev_2009}. More specifically, they arise from a flat constant background, reach a transient high amplitude, and then disappear back into the same flat background. Such solutions were first reported for the nonlinear Schr\"{o}dinger (NLS) equation by Peregrine in 1983 \cite{Peregrine}. In recent years, such waves were linked to freak waves in the ocean \cite{Ocean_rogue_review,Pelinovsky_book} and extreme events in optics \cite{Solli_Nature,Wabnitz_book}, and were observed in controlled experiments in water tanks \cite{Tank1,Tank2} and optical fibers \cite{Fiber1,Fiber2,Fiber3}. Motivated by these physical applications, rogue waves have been derived in a large number of integrable nonlinear wave equations such as the NLS equation \cite{AAS2009,DGKM2010,ACA2010,DPMB2011,KAAN2011,GLML2012,OhtaJY2012,DPMVB2013}, the derivative NLS equation \cite{XuHW2011,GLML2013}, the three-wave interaction equation \cite{BCDL2013}, the Davey-Stewartson equations \cite{OhtaJKY2012,OhtaJKY2013}, and many others \cite{AANJM2010,OhtaJKY2014,ASAN2010,TaoHe2012,BDCW2012,ManakovDark,Chow,PSLM2013,Grimshaw_rogue,MuQin2016,LLMSasa2016,LLMFZ2016,Degasperis1,Degasperis2}. Indeed, rogue waves are caused by baseband modulation instability of the constant background \cite{ManakovDark}. Thus, any integrable equation with baseband modulation instability is expected to admit rogue waves which can be derived by integrable techniques.

A notable integrable equation that admits rogue waves is the nonlocal NLS equation
\[ \label{eq1:PTNLS}
\textrm{i} u_t(x,t)=u_{xx}(x,t)+2 u^2(x,t)u^*(-x,t),
\]
which was introduced by Ablowitz and Musslimani in 2013 \cite{Ablowitz_PRL}. The nonlinearity in this equation is nonlocal and parity-time ($\cal{PT}$) symmetric \cite{Yang_PTreview}, which differs from all previous integrable equations. A potential physical application of this equation has also been identified in the context of an unconventional system of magnetics \cite{magnetics}. Since its introduction, this nonlocal NLS equation has been heavily studied \cite{Ablowitz_nonlinearity,WYY2016,HXLM2016,Gerdjikov2017,YangPTNLSsoliton,FengPTNLSsoliton,YangYang2018}. In particular, its rogue waves were derived by Darboux transformation in \cite{YangYang2018}, where three families of rogue waves with polynomial degrees $N(N+1)$, $N(N-1)+1$ and $N^2$ were reported ($N$ here is an arbitrary positive integer). Recalling that the local NLS equation admits only a single family of rogue waves with polynomial degrees $N(N+1)$ \cite{ACA2010,OhtaJY2012}, it is then clear that the nonlocal NLS equation (\ref{eq1:PTNLS}) admits a much wider variety of rogue waves. However, an important question which was not clarified in \cite{YangYang2018} is whether there exist additional families of rogue waves in the nonlocal equation (\ref{eq1:PTNLS}). This question will be addressed in this article.

In this paper, we derive rogue waves in the nonlocal NLS equation (\ref{eq1:PTNLS}) by the bilinear Kadomtsev-Petviashvili (KP) reduction method. Compared to the previous KP-reduction method for rogue waves in the local NLS equation \cite{OhtaJY2012}, a key difference in our current method is the realization that the bilinear equations for the nonlocal NLS equation admit a Gram-determinant solution which has a more general structure than the one reported in \cite{OhtaJY2012}. This realization allows us to derive broader classes of rogue waves with any polynomial degree of the form $[(N_{1}-N_{2})^2+(N_{1}-N_{2})+(M_{1}-M_{2})^2+(M_{1}-M_{2})]/2$, where $N_1, N_2, M_1$ and $M_2$ are arbitrary non-negative integers under the constraint of $N_{1}+N_{2}=M_{1}+M_{2}$. For special choices of the $(N_1, N_2, M_1, M_2)$ values, these rogue waves reproduce the previous three solution families reported in \cite{YangYang2018}. But they also contain many new rogue-wave families and are thus more general. Explicit expressions of our rogue waves are given as Gram determinants with Schur-polynomial matrix elements, and rogue waves from the new solution families are shown to feature distinctive patterns. In addition, we conjecture that the rogue waves derived in this article are all the rogue waves admitted by the nonlocal NLS equation (\ref{eq1:PTNLS}).

\section{General rogue wave solutions}
Without loss of generality, we consider rogue waves in the nonlocal NLS equation (\ref{eq1:PTNLS}) which approach unit-amplitude background at large $x$ and $t$,
\[\label{BoundaryCond1}
u(x,t) \rightarrow e^{-2\textrm{i}t}, \ \ x, t \rightarrow \pm \infty.
\]

Before presenting rogue wave solutions in Eq. (\ref{eq1:PTNLS}), we first introduce elementary Schur polynomials $S_k(\mbox{\boldmath $x$})$
which are defined via the generating function,
\[ \label{def_Schur}
\sum_{k=0}^{\infty}S_k(\mbox{\boldmath $x$})\lambda^k
=\exp\left(\sum_{j=1}^{\infty}x_j\lambda^j\right),
\]
where $\mbox{\boldmath $x$}=(x_1,x_2,\cdots)$. Specifically, we have
\[
S_0(\mbox{\boldmath $x$})=1, \quad S_1(\mbox{\boldmath $x$})=x_1,
\quad S_2(\mbox{\boldmath $x$})=\frac{1}{2}x_1^2+x_2, \quad \cdots, \quad
S_{k}(\mbox{\boldmath $x$}) =\sum_{l_{1}+2l_{2}+\cdots+ml_{m}=k} \left( \ \prod _{j=1}^{m} \frac{x_{j}^{l_{j}}}{l_{j}!}\right).
\]
Now we present rogue waves in terms of these Schur polynomials.

\textbf{Theorem 1.} \emph{The $\cal{PT}$-symmetric nonlinear Schr\"{o}dinger equation (\ref{eq1:PTNLS}) under boundary conditions (\ref{BoundaryCond1}) admits the following rational rogue-wave solutions}
\[ \label{Th1-solutions}
u(x,t)=e^{-2 \textrm{i} t}\frac{\sigma_{1}}{\sigma_{0}},
\]
\emph{where}
\[ \label{Sigma-n}
\sigma_{n}=
\left|
\begin{array}{cc}
  \Gamma_{1,1}^{(n)} & \Gamma_{1,2}^{(n)} \\
  \Gamma_{2,1}^{(n)} & \Gamma_{2,2}^{(n)}
\end{array}
\right|,
\]
\emph{$\Gamma_{i,j}^{(n)}$ are $N_{i} \times M_{j}$ block matrices defined by}
\[\label{Blockmatrix}
\Gamma_{i, j}^{(n)}=
\left(
m_{2k-i, \, 2l-j}^{(n)}
\right)_{1\leq k \leq N_{i}, \, 1\leq l \leq M_{j}},
\]
\emph{$N_1, N_2, M_1$ and $M_2$ are arbitrary non-negative integers with the constraint of $N_{1}+N_{2}=M_{1}+M_{2}$, the matrix elements in $\Gamma_{i, j}^{(n)}$ are defined by}
\[ \label{matrixmnij}
m_{i,j}^{(n)}=\sum_{\nu=0}^{\min(i,j)} \Phi_{i,\nu}^{(n)}\Psi_{j,\nu}^{(n)},
\]
\emph{with}
\begin{eqnarray}
\Phi_{i,\nu}^{(n)}=\frac{1}{2^{\nu}}\sum_{k=0}^{i-\nu}a_{k}S_{i-\nu-k}(\textbf{\emph{x}}^{+}(n)+\nu \textbf{\emph{s}}), \quad   \Psi_{j,\nu}^{(n)}=\frac{1}{2^{\nu}}\sum_{l=0}^{j-\nu}b_{l}S_{j-\nu-l}(\textbf{\emph{x}}^{-}(n)+\nu \textbf{\emph{s}}),\label{phipsiexp}
\end{eqnarray}
vectors $\textbf{\emph{x}}^{\pm}(n)=\left(  x_{1}^{\pm}(n), x_{2}^{\pm}, x_{3}^{\pm},\cdots \right)$ \emph{and} $\emph{\textbf{s}}=(s_{1}, s_{2}, \cdots)$ \emph{are defined by}
\begin{eqnarray}
&& x_{1}^{\pm}(n)=x\mp 2 \textrm{i} t\pm n-\frac{1}{2}, \quad x_{k}^{\pm}=\frac{x\mp2^{k} \textrm{i} t}{k!}-r_{k},\ (k>1),  \label{skrkexpcoeff} \\
&& \sum_{k=1}^{\infty} r_{k}\lambda^{k}=\ln\left( \cosh\frac{\lambda}{2} \right),  \quad
\sum_{k=1}^{\infty} s_{k}\lambda^{k}=\ln\left( \frac{2}{\lambda} \tanh\frac{\lambda}{2} \right),  \label{skrkexpcoeff2}
\end{eqnarray}
\emph{$a_{k}$, $b_{l}$ are complex constants which are given as}
\[
a_0=b_0=1, \quad a_2=a_4=\cdots=a_{even}=0, \quad b_2=b_4=\cdots=b_{even}=0,
\]
\[ \label{a_recur}
\Re(a_{2k-1})=\frac{1}{2(2k-1)!}-\sum_{j=1}^{k-1} \frac{\Re(a_{2j-1})}{(2k-2j)!}+\sum_{i=1}^{k-1}\sum_{j=1}^{k-i}\frac{a_{2i-1}a_{2j-1}^*}{2(2k-2i-2j+1)!}, \quad k=1, 2, \cdots,
\]
\[ \label{b_recur}
\Re(b_{2k-1})=\frac{1}{2(2k-1)!}-\sum_{j=1}^{k-1} \frac{\Re(b_{2j-1})}{(2k-2j)!}+\sum_{i=1}^{k-1}\sum_{j=1}^{k-i}\frac{b_{2i-1}b_{2j-1}^*}{2(2k-2i-2j+1)!}, \quad k=1, 2, \cdots,
\]
\emph{and $\Im(a_{2k-1})$, $\Im(b_{2k-1})$ $(k=1, 2, \cdots)$ are free parameters. Here, $\Re$ and $\Im$ represent the real and imaginary parts of a complex number, and the asterisk $*$ represents complex conjugation.}

The degrees of polynomials $\sigma_{n}(x,t)$ in Theorem 1 and their leading terms are given by the following theorem.

\textbf{Theorem 2.}  \emph{The degrees of polynomials $\sigma_{n}(x,t)$ in both $x$ and $t$ for rogue waves in Theorem 1 are}
\[ \label{polydegree}
\deg(\sigma_{n})=\frac{1}{2}\left[(N_{1}-N_{2})^2+(N_{1}-N_{2})+(M_{1}-M_{2})^2+(M_{1}-M_{2})\right].
\]
\emph{In addition, the leading (highest-power) terms in $\sigma_{n}(x,t)$ are}
\begin{eqnarray}\label{LeadingTerm}
\sigma_{n}=c_{0}(x-2\textrm{i}t)^{(N_{1}-N_{2})(N_{1}-N_{2}+1)/2}(x+2\textrm{i}t)^{(M_{1}-M_{2})(M_{1}-M_{2}+1)/2}+ (\text{lower degree terms}),
\end{eqnarray}
\emph{where $c_0$ is a $\left[N_{1}, N_{2}, M_{1}, M_{2}\right]$-dependent but $n$-independent constant.}

\textbf{Remark 1.} In Theorem 1, $\Re(a_{2k-1})$ and $\Re(b_{2k-1})$ are given by recursive relations (\ref{a_recur})-(\ref{b_recur}), and the free parameters in rogue waves (\ref{Th1-solutions}) are
\begin{eqnarray*}
&& \Im(a_1), \Im(a_3), \Im(a_5), \cdots; \hspace{0.1cm} \Im(b_1), \Im(b_3), \Im(b_5), \cdots,
\end{eqnarray*}
where the index $k$ for $a_k$ does not exceed max$(2N_1-1, 2N_2-2)$, and the index $k$ for $b_k$ does not exceed max$(2M_1-1, 2M_2-2)$.
In Appendix A, we will show that when $N_{2}<N_{1}$, parameters $\Im(a_{2(N_{1}-N_{2})+1})$, $\Im(a_{2(N_{1}-N_{2})+3})$, $\cdots$ drop out (i.e., they do not affect $\sigma_n$'s determinant values); when $N_{2}=N_{1}$ or $N_{2}=N_{1}+1$, all $\Im(a_1)$, $\Im(a_3)$, $\cdots$ drop out; and when $N_2>N_1+1$, parameters $\Im(a_{2(N_{2}-N_{1})-1})$, $\Im(a_{2(N_{2}-N_{1})+1})$, $\cdots$ drop out. Likewise, when $M_{2}<M_{1}$, parameters $\Im(b_{2(M_{1}-M_{2})+1})$, $\Im(b_{2(M_{1}-M_{2})+3})$, $\cdots$ drop out; when $M_{2}=M_{1}$ or $M_{2}=M_{1}+1$, all $\Im(b_1)$, $\Im(b_3)$, $\cdots$ drop out; and when $M_2>M_1+1$, parameters $\Im(b_{2(M_{2}-M_{1})-1})$, $\Im(b_{2(M_{2}-M_{1})+1})$, $\cdots$ drop out.
Furthermore, since the nonlocal NLS equation (\ref{eq1:PTNLS}) is time-translation invariant, by a shift of the $t$ axis, we can remove one more real parameter. Thus, the number of irreducible free real parameters in the rogue-wave solutions of Theorem 1 is given in the following table:
\begin{table}[h]
\caption{Number of irreducible free real parameters in rogue waves of Theorem 1}
\begin{center}
  \begin{tabular}{ | c | c | c|  c | c| }  \hline
                   & $M_2\le M_1$        & $M_2>M_1$ \\ \hline
  $N_{2}\le N_{1}$ & $N_1-N_2+M_1-M_2-1$   & $N_1-N_2+M_2-M_1-2$    \\ \hline
  $N_2>N_1$        & $N_2-N_1+M_1-M_2-2$ & $N_2-N_1+M_2-M_1-3$  \\ \hline
  \end{tabular}
\end{center}
\end{table}

\textbf{Remark 2.} Compared to the $\sigma_n$ function in Ref.~\cite{OhtaJY2012} for rogue waves in the local NLS equation, the present $\sigma_n$ function in Eq. (\ref{Sigma-n}) has a more general matrix structure. Indeed, the $\sigma_n$ function in Ref.~\cite{OhtaJY2012} only corresponds to the subblock $\Gamma_{1,1}^{(n)}$ in our current $\sigma_n$ function (\ref{Sigma-n}). This more general matrix structure (\ref{Sigma-n}) for $\sigma_n$ is one of our key realizations, and it leads to a much wider variety of rogue waves in the nonlocal NLS equation than in its local counterpart (see Theorem 2 and Sec. 3 below).

\textbf{Remark 3.} Theorem 2 shows that the degrees of polynomials in rogue waves of the nonlocal NLS equation are much richer than those in the local NLS equation. Indeed, when $N_1=M_1=N$ and $N_2=M_2=0$, the above formula gives a polynomial degree of $N(N+1)$, which reproduces the polynomial degree of rogue waves in the local NLS equation \cite{ACA2010,OhtaJY2012}. But other choices of the $(N_1, N_2, M_1, M_2)$ values would give rogue waves with polynomial degrees beyond $N(N+1)$. The above polynomial-degree formulas are also much broader than those in the three types of rogue waves reported for the nonlocal NLS equation in \cite{YangYang2018}, which we will elaborate in Sec. 3.

\textbf{Remark 4.} Theorem 2 also shows that when $N_{2}=N_{1}$ or $N_{2}=N_{1}+1$, and $M_{2}=M_{1}$ or $M_{2}=M_{1}+1$, $\deg(\sigma_{n})=0$. In these cases, $\sigma_n$ is a constant, and thus the solution $u(x,t)$ in Eq. (\ref{Th1-solutions}) is also a constant, not a true rogue wave.

Proofs of these two theorems will be given in Sec. 4.

\section{Analysis of rogue waves}
Rogue waves in Theorem 1 contain a wide variety of solutions in the nonlocal NLS equation (\ref{eq1:PTNLS}). In particular, they
contain the three families of rogue waves previously reported in \cite{YangYang2018} as special cases. In addition, they contain new rogue-wave families which have not been reported before.

\subsection{Previous rogue waves as special cases of current solutions} \label{s:31}
In this subsection, we show that rogue waves in Theorem 1 contain the three families of rogue waves previously derived in \cite{YangYang2018} by Darboux transformation. Since the present rogue waves and those in \cite{YangYang2018} are derived by different methods and expressed in different ways, a direct proof of this statement is cumbersome. Thus, we will argue differently below.

First, we take $\left[ N_{1},  N_{2}, M_{1},  M_{2}\right]= \left[N, 0, N, 0\right]$ in Theorem 1. In this case, the resulting rogue waves have a polynomial degree of $N(N+1)$ (see Theorem 2), which matches type-I rogue waves in \cite{YangYang2018}. In addition, the number of irreducible free real parameters in these rogue waves is $2N-1$ (see Remark 1), which also matches that number in type-I rogue waves in \cite{YangYang2018} (see Remark 1 in that article). Furthermore, when $N=1$, the explicit expression of rogue waves in Theorem 1 is
\begin{eqnarray}\label{QuadraticRws1}
u(x,t)=e^{-2\textrm{i}t}\left(1 + \frac{4(4  \textrm{i} \hat{t}-1)}{ 16 \hat{t}^{\hspace{0.03cm}2} + 4(x+\textrm{i}x_{0})^2+1}\right),
\end{eqnarray}
where $\hat{t}=t-\left[\Im(a_1)-\Im(b_1)\right]/4$, $x_{0}=\left[\Im(a_1)+\Im(b_1)\right]/2$, and $\Im(a_1)$, $\Im(b_1)$ are free real parameters. This solution matches the type-I rogue waves with $N=1$ in \cite{YangYang2018} [see Eq. (79) there]. When $N=2$, we have compared the explicit expressions of the present solution with the type-I rogue waves with $N=2$ in \cite{YangYang2018} and found them equivalent as well under the parameter mappings of
\begin{equation*}
s_0 \to \Im(a_1), \hspace{0.2cm}  r_0 \to \Im(b_1), \hspace{0.2cm}  s_1\to \frac{1}{6} \left(-4 (\Im(a_1))^3+\Im(a_1)+12 \Im(a_3)\right), \hspace{0.2cm} r_1\to \frac{1}{6} \left(-4 (\Im(b_1))^3+\Im(b_1)+12 \Im(b_3)\right),
\end{equation*}
where $(s_0, r_0, s_1, r_1)$ are parameters in type-I rogue waves of \cite{YangYang2018}. From these, we can conclude that rogue waves in Theorem 1 with parameter choices of $N_1=M_1$ and $N_2=M_2=0$ reproduce the type-I rogue waves in \cite{YangYang2018}.

Second, we take $\left[ N_{1},  N_{2}, M_{1},  M_{2}\right]= \left[N, 0, N-1, 1\right]$ in Theorem 1. In this case, the resulting rogue waves have a polynomial degree of $N(N-1)+1$, and their number of irreducible free real parameters is zero when $N=1$ and $2N-3$ when $N>1$. These numbers match those of type-II rogue waves in \cite{YangYang2018}. In addition, when $N=1$, the rogue wave of Theorem 1 can be simplified as
\begin{eqnarray}\label{LinearRws2}
u_{1}(x,t)=e^{-2\textrm{i}t}\left(1+\frac{1 }{x-2  \textrm{i}  \hat{t}}\right),
\end{eqnarray}
where $\hat{t}=t-\Im(a_1)/2$. This solution matches the type-II rogue wave with $N=1$ in \cite{YangYang2018} [see Eq. (80) there]. When $N=2$, the rogue wave of Theorem 1 can be simplified as
\[\label{cubicsolution}
u(x,t)=e^{-2\textrm{i}t}\left(1+\frac{3\left(2 x-4\textrm{i}\hat{t} +1\right)^2}{4\left(x -2\textrm{i}\hat{t}\ \right)^3 - 3\left(x - 6 \textrm{i} \hat{t}+ \textrm{i} \beta \right)}\right),
\]
where $\hat{t}=t-\Im(a_1)/2$, and $\beta= 4 \left(\Im(a_1)\right)^3+ 5 \Im(a_1)- 12\Im(a_3)$. This solution matches the type-II rogue waves with $N=2$ in \cite{YangYang2018} as well [see Eq. (82) there]. From these, we can conclude that rogue waves in Theorem 1 with parameter choices of $\left[ N_{1},  N_{2}, M_{1},  M_{2}\right]= \left[N, 0, N-1, 1\right]$ reproduce the type-II rogue waves in \cite{YangYang2018}.

Thirdly, we take $\left[ N_{1},  N_{2}, M_{1},  M_{2}\right]= \left[N, 0, 0, N\right]$ in Theorem 1. In this case, the resulting rogue waves have a polynomial degree of $N^2$, and their number of irreducible free real parameters is $2N-2$. These numbers match those of type-III rogue waves in \cite{YangYang2018} (the number of free real parameters in type-III rogue waves was quoted as $2N$ in Remark 1 of \cite{YangYang2018}, leading to $2N-1$ irreducible free real parameters after a time shift, which is incorrect). When $N=1$, the rogue wave of Theorem 1 turns out to be the same as that in Eq. (\ref{LinearRws2}), which matches the type-III rogue wave with $N=1$ in \cite{YangYang2018}. When $N=2$, we have verified that the rogue wave from Theorem 1 is equivalent to the type-III rogue wave with $N=2$ in \cite{YangYang2018} as well under the parameter mappings of
\begin{equation*}
s_0 \to \Im(a_1), \hspace{0.2cm} r_0 \to \Im(b_1), \hspace{0.2cm} s_1\to \frac{1}{6} \left(-4 (\Im(a_1))^3+\Im(a_1)+12 \Im(a_3)\right).
\end{equation*}
From these, we can conclude that rogue waves in Theorem 1 with parameter choices of $\left[ N_{1},  N_{2}, M_{1},  M_{2}\right]= \left[N, 0, 0, N\right]$ reproduce the type-III rogue waves in \cite{YangYang2018}.

\subsection{New families of rogue waves}
Rogue waves in Theorem 1 contain not only those which have been reported before, but also many new solutions. These new solutions are reflected in two ways. One is that rogue waves in Theorem 1 admit new polynomial degrees beyond those reported in \cite{YangYang2018}. For instance, the current rogue waves admit polynomial degrees such as 10 and 11, which are not possible in rogue waves reported in \cite{YangYang2018}. The other is that, even for the same polynomial degrees as those contained in \cite{YangYang2018}, Theorem 1 could admit rogue waves which are different from those in \cite{YangYang2018}. In this subsection, we discuss and illustrate such new solutions.

First, we consider rogue waves in Theorem 1 which feature new polynomial degrees. As examples, we consider two choices of
$\left[ N_{1},  N_{2}, M_{1},  M_{2}\right]= \left[4, 0, 2, 2\right]$ and $\left[4, 0, 1, 3\right]$. For these choices, the polynomial degrees of rogue waves can be derived from Theorem 2 as 10 and 11 respectively. It is easy to see that these polynomial degrees are not possible in previous rogue waves reported in \cite{YangYang2018}, whose degrees are only of types $N(N+1)$, $N(N-1)+1$ and $N^2$; thus they are new rogue wave solutions in the nonlocal NLS equation (\ref{eq1:PTNLS}). The degree-10 rogue waves contain four free real parameters $\Im(a_{1}), \Im(a_{3}), \Im(a_{5})$ and $\Im(a_{7})$, and
the degree-11 rogue waves contain five free real parameters $\Im(a_{1}), \Im(a_{3}), \Im(a_{5})$, $\Im(a_{7})$ and $\Im(b_{1})$ (both without parameter reduction by time shifting). To illustrate, two of such solutions (one for each degree) are displayed in Fig. 1. It is seen that these rogue waves exhibit 10 and 11 singular (blowup) points in the $(x, t)$ plane respectively, and these singularities form new patterns which have not been seen before.

\begin{figure}[htb]
   \begin{center}
   \vspace{-2cm}
   \includegraphics[scale=0.385, bb=0 0 385 567]{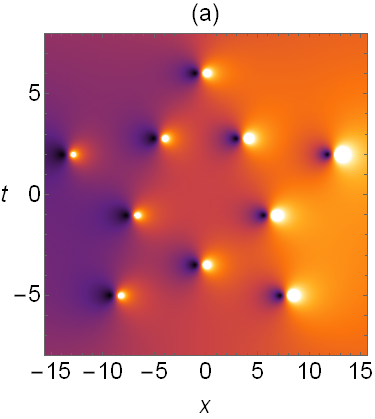} \hspace{0.5cm}
     \includegraphics[scale=0.385, bb=0 0 385 567]{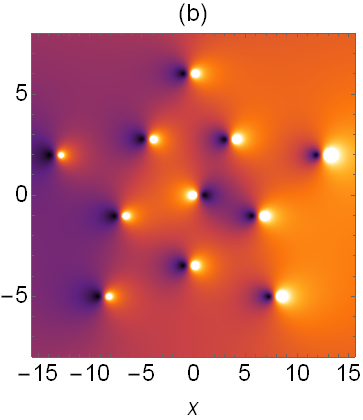}
   \caption{(a) A degree-10 rogue wave with parameters $\Im(a_{1})=\Im(a_{3})=0$, $\Im(a_{5})=1000$ and $\Im(a_{7})=-250$.
   (b) A degree-11 rogue wave with parameters $\Im(a_{1})=\Im(b_{1})=\Im(a_{3})=0$, $\Im(a_{5})=1000$ and $\Im(a_{7})=-250$.}
   \end{center}
\end{figure}

Next, we show that Theorem 1 admits rogue waves which are new solutions even though their polynomial degrees are not new. As an example, we consider the parameter choice of $\left[ N_{1},  N_{2}, M_{1},  M_{2}\right]= \left[0, 4, 2, 2\right]$, which gives rogue waves with a polynomial degree of 6. This polynomial degree is attainable from type-I rogue waves in \cite{YangYang2018} with $N=2$ (which is equivalent to rogue waves in Theorem 1 with the choice of $\left[ N_{1},  N_{2}, M_{1},  M_{2}\right]= \left[N, 0, N, 0\right]$, see the previous subsection). Thus, this polynomial degree is not new. But we will show that these rogue waves do not belong to type-I rogue waves in \cite{YangYang2018} and are thus true new solutions.

For this choice of $\left[ N_{1},  N_{2}, M_{1},  M_{2}\right]$ values, the corresponding rogue waves contain three free real parameters $\Im(a_{1}), \Im(a_{3})$ and $\Im(a_{5})$. After a shift of time, $\Im(a_{1})$ can be further removed. Thus, we can set $\Im(a_{1})=0$ without loss of generality. Then, denoting $\zeta=x-2\textrm{i}t$, the solution expression of these rogue waves is
\[\label{degreesixrws1}
u(x,t)=e^{-2\textrm{i}t}\frac{\sigma_{1}}{\sigma_{0}},
\]
where
\begin{eqnarray}
&&\sigma_{0}=4\zeta^6 - 15 \zeta^3(3\zeta-2x)- 60 \textrm{i}\Im(a_{3})  \zeta^3+ 180\textrm{i}\Im(a_{5})\zeta +180\left[t-\Im(a_{3})\right]^2,   \label{sigma0A} \\
&&\sigma_{1}= \sigma_{0} + 24\zeta^5 + 60 \zeta^4+30 \zeta^2 (3x-2\zeta)-180 \textrm{i} \left[ \Im(a_{3})\zeta(\zeta+1)-(t)\zeta-\Im(a_{5}) \right].  \label{sigma1A}
\end{eqnarray}
For three choices of $\Im(a_{3})$ and $\Im(a_{5})$ values, the corresponding rogue waves are displayed in Fig. 2. It is seen that these solutions feature six singularities in the $(x,t)$ plane, and these singularities form interesting patterns such as a triangle in panel (a), a pentagon in panel (b), and an intermediate state in panel (c). None of these patterns has been reported before.

\begin{figure}[htb]
   \begin{center}
   \vspace{-1.5cm}
   \includegraphics[scale=0.360, bb=0 0 385 567]{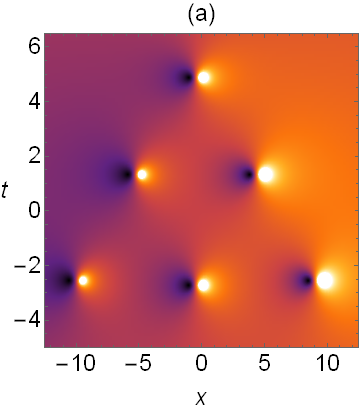}   \hspace{0.25cm}
   \includegraphics[scale=0.350, bb=0 0 385 567]{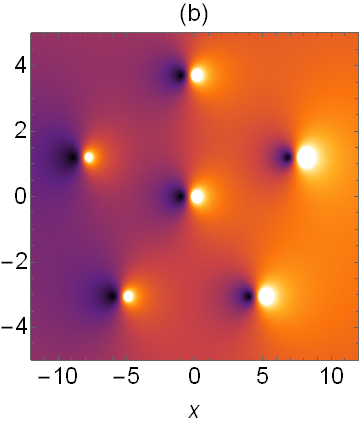} \hspace{0.25cm}
   \includegraphics[scale=0.350, bb=0 0 385 567]{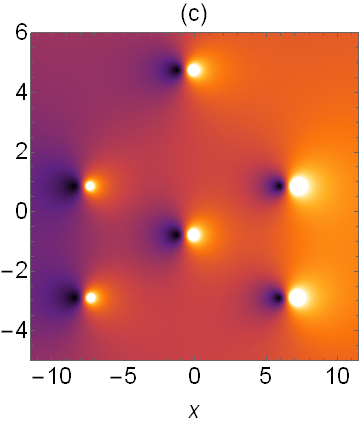}
   \caption{Three new rogue waves with polynomial degree six, which are obtained from Theorem 1 under $\left[ N_{1},  N_{2}, M_{1},  M_{2}\right]= \left[0, 4, 2, 2\right]$. Free parameters in these solutions are chosen as (a) $\Im(a_{1})=0$, $\Im(a_{3})=60$, $\Im(a_{5})=0$; (b) $\Im(a_{1})=\Im(a_{3})=0$, $\Im(a_{5})=600$; (c) $\Im(a_{1})=0$, $\Im(a_{3})=35$, $\Im(a_{5})=800$.}
   \end{center}
\end{figure}

To show these degree-six rogue waves are new solutions, we compare them with degree-six rogue waves of type-I reported in \cite{YangYang2018}. Those rogue waves are
\[\label{degreesixrws2}
u(x,t)=e^{-2\textrm{i}t}\frac{\sigma_{1}}{\sigma_{0}},
\]
with the leading-order terms of $\sigma_n$ given as
\begin{eqnarray}\label{LeadingTerm2}
\sigma_{n}=64 (x^2+4t^2)^3 + (\text{lower degree terms}).
\end{eqnarray}
By comparing these leading-order terms of $\sigma_n$ with those in Eqs. (\ref{sigma0A})-(\ref{sigma1A}), we see that they are clearly very different. Thus, these two types of degree-six rogue waves cannot be equivalent to each other. Indeed, solution patterns in Fig. 2 for the present degree-six rogue waves in Eq. (\ref{degreesixrws1}) are very different from those for degree-six rogue waves of type-I as reported in \cite{YangYang2018}. Thus, there is no question that the present degree-six rogue waves (\ref{degreesixrws1}) are new solutions to the nonlocal NLS equation (\ref{eq1:PTNLS}).

\textbf{Remark 5.} We should point out that, while Theorem 1 gives a wide variety of rogue wave families characterized by their choices of $\left[ N_{1},  N_{2}, M_{1},  M_{2}\right]$ values, some of those solution families with different $\left[ N_{1},  N_{2}, M_{1},  M_{2}\right]$ values may be equivalent to each other. For example,
\begin{enumerate}
\item solutions with $\left[ N_{1},  N_{2}, M_{1},  M_{2}\right]= \left[\widetilde{N}_{1}, \widetilde{N}_{2}, \widetilde{M}_{1},\widetilde{M}_{2}\right]$ and those with $\left[\widetilde{N}_{2}, \widetilde{N}_{1}+1, \widetilde{M}_{2}, \widetilde{M}_{1}+1\right]$ are equivalent to each other;
\item as a special case of the above equivalency, solutions with $\left[ N_{1},  N_{2}, M_{1},  M_{2}\right]=[N, 0, N, 0]$ and those with $[0, N+1, 0, N+1]$ are equivalent to each other;
\item as a corollary of the above equivalency, solutions with $\left[ N_{1},  N_{2}, M_{1},  M_{2}\right]=[N, K, N, K]$ are equivalent to those with $[N-K, 0, N-K, 0]$ when $N>K$ and equivalent to those with $[0, K-N, 0, K-N]$ when $N<K$.
\end{enumerate}
These equivalencies will be proved in Appendix B. Note that two solutions are equivalent to each other not only when they can be made equal to each other, but also when they can be related as $u(x, t)$ and $u^*(-x, -t)$. The reason is that the nonlocal NLS equation (\ref{eq1:PTNLS}) is \PT-symmetric --- thus if $u(x,t)$ is a solution, so is $u^*(-x, -t)$. A quick way to identify equivalent solutions with different $\left[ N_{1},  N_{2}, M_{1},  M_{2}\right]$ values is to compare their polynomial degrees and leading-order terms according to Theorem 2, and compare their numbers of irreducible free real parameters according to Table 1. If their polynomial degrees and numbers of free real parameters match each other, and their leading-order polynomial terms are the same or related as $u(x, t)$ and $u^*(-x, -t)$, then it is almost certain that solutions with such different $\left[ N_{1},  N_{2}, M_{1},  M_{2}\right]$ values would be equivalent to each other.

\section{Derivation of rogue waves}
In this section, we derive the rogue wave solutions given in Theorem 1. This derivation uses the bilinear KP-reduction method \cite{Hirota}. This method has been applied for the derivation of rogue waves in the local NLS equation \cite{OhtaJY2012} and several other $(1+1)$- and $(2+1)$-dimensional integrable equations before \cite{OhtaJKY2012,OhtaJKY2013,OhtaJKY2014,Chen_Juntao,XiaoeYong2018,SunLian2018}. Compared to earlier applications of this method on $(1+1)$-dimensional systems, a key new feature in our method is that the Gram determinants in our solutions have a structure which is more general. This is the reason why rogue waves in Theorem 1 feature a wider variety of solution families with richer polynomial degrees than those in the local NLS equation and other $(1+1)$-dimensional integrable equations \cite{OhtaJY2012,OhtaJKY2014,Chen_Juntao,XiaoeYong2018}.

First, via the variable transformation
\[\label{usolution}
u=e^{-2\textrm{i}t}\frac{g}{f},
\]
the $\cal{PT}$-symmetric NLS equation (\ref{eq1:PTNLS}) is transformed into the bilinear form,
\begin{eqnarray}\label{Bilineform11}
 \left.
   \begin{array}{ll}
    (D_{x}^{2} +2 ) f \cdot f = 2g \bar{g},\\
   (D_{x}^{2} -\textrm{i} D_{t}) g \cdot f =0,
   \end{array}
 \right\}
\end{eqnarray}
where the overbar $\bar{\ }$ on a function $g(x,t)$ is defined as
\[
\bar{g}(x,t)\equiv g^*(-x,t),
\]
and $f$ is a complex function satisfying the condition
\[
\bar{f}(x,t)=f(x,t).
\]
Here, $D$ is Hirota's bilinear differential operator defined by
\begin{eqnarray*}
&&P \left(D_{x}, D_{y}, D_{t},\cdots\right) F(x,y,t,\cdots) \cdot G(x,y,t,\cdots) \\
&& = P\left(\partial_{x}-\partial_{x'}, \partial_{y}-\partial_{y'}, \partial_{t}-\partial_{t'}, \cdots \right) G(x',y',t',\cdots)|_{x'=x, y'=y, t'=t,\cdots},
\end{eqnarray*}
where $P$ is a polynomial of $D_{x}$, $D_{y}$, $D_{t}, \cdots .$

In order to derive solutions to the bilinear equations (\ref{Bilineform11}), we consider a $(2+1)$-dimensional bilinear system,
\begin{eqnarray}\label{Bilineform12}
 \left.
   \begin{array}{ll}
     (D_{x}D_{y} +2 ) f \cdot f = 2g h,\\
     (D_{x}^{2} -\textrm{i} D_{t}) g \cdot f =0,
   \end{array}
 \right\}
\end{eqnarray}
where $h$ is another complex function. Algebraic solutions of Gram-determinant types for this higher-dimensional bilinear system have been given in \cite{OhtaJY2012}. By choosing Gram determinants which are more general than previously used, we can show these algebraic solutions satisfy the dimension reduction condition
\[\label{DimentionPara}
\left(\partial_{x}-\partial_{y}\right) f=Cf,
\]
where $C$ is some constant. Thus, these solutions satisfy the $(1+1)$-dimensional bilinear system
\begin{eqnarray}\label{Bilineform13}
 \left.
   \begin{array}{ll}
     (D_{x}^2 +2 ) f \cdot f = 2g h,\\
     (D_{x}^{2} -i D_{t}) g \cdot f =0.
   \end{array}
 \right\}
\end{eqnarray}
By judiciously choosing parameters in these Gram determinants, we can further show that they would satisfy the nonlocal reduction condition
\[\label{NonlocPara}
\bar{f}=f,\ \ \ h=\bar{g}.
\]
Thus, these solutions also satisfy the original bilinear equations (\ref{Bilineform11}). Details of these steps are given next.

\subsection{Algebraic solutions for the $(1+1)$-dimensional bilinear system (\ref{Bilineform13})}
Introducing Gram determinants
\begin{eqnarray}\label{001}
\tau_{n}=\det_{1 \leq i,j \leq N} \left( m_{i,j}^{(n)} \right),
\end{eqnarray}
where
\begin{eqnarray}
&&m_{i,j}^{(n)}= \mathcal{A}_{i} \mathcal{B}_{j} \tilde{m}^{(n)}|_{p=1,q=1}, \quad \tilde{m}^{(n)}=\frac{1}{p+q}\left(-\frac{p}{q}\right)^{n}e^{\tilde{\xi}+\tilde{\eta}}, \label{002-1} \\
&& \tilde{\xi}=\frac{1}{p}x_{-1}+ p x_{1}+p^2x_{2}, \quad \tilde{\eta}=\frac{1}{q}x_{-1}+ q x_{1}-q^2x_{2}, \label{002-2}
\end{eqnarray}
$\mathcal{A}_{i}$ and $\mathcal{B}_{j}$ are differential operators with respect to $p$ and $q$ as
\begin{eqnarray}\label{003}
\left.
  \begin{array}{ll}
    \mathcal{A}_{i}=\sum_{k=0}^{i}\frac{a_{k}}{(i-k)!}\left(p\partial_{p}\right)^{i-k} \vspace{0.25cm}\\
 \mathcal{B}_{j}=\sum_{l=0}^{j}\frac{b_{l}}{(j-l)!}\left(q\partial_{q}\right)^{j-l}
  \end{array}
\right\},
\end{eqnarray}
and $a_{k}$, $b_{k}$  are complex constants, then it was shown in Ref. \cite{OhtaJY2012} that for arbitrary sequences of indices
\[ \label{arbiindex}
\left(i_{1}, i_{2},..., i_{N}; j_{1}, j_{2},...,j_{N}\right),
\]
the determinant
\begin{eqnarray}\label{dimentionalindex}
 \tau_{n}=\det_{1\leq \nu,\mu\leq N}\left( m_{i_{\nu}, j_{\mu}}^{(n)} \right)
\end{eqnarray}
satisfies the following bilinear equations
\begin{eqnarray}\label{004}
\left.
\begin{array}{c}
   \left(D_{x_{1}}D_{x_{-1}}-2\right)\tau_{n}\cdot\tau_{n}=-2 \tau_{n+1} \tau_{n-1} \\
  \left(D_{x_{1}}^2-D_{x_{2}}\right)\tau_{n+1} \cdot \tau_{n}=0.
\end{array}
\right\}
\end{eqnarray}

In Ref.~\cite{OhtaJY2012}, the authors chose the determinant as
\begin{eqnarray}\label{005}
\tau_{n}^{(1)}=\det_{1\leq i,j \leq N}\left( m_{2i-1, 2j-1}^{(n)} \right),
\end{eqnarray}
and showed that this determinant satisfies the dimension reduction condition
\begin{eqnarray}\label{007}
(\partial_{x_{1}}+\partial_{x_{-1}}) \tau_{n}^{(1)}=4N \tau_{n}^{(1)}.
\end{eqnarray}

In this paper, we show that in addition to the determinant (\ref{005}), there exists other types of determinants $\tau_{n}$ which also satisfy the dimension reduction condition (\ref{007}). Specifically, any one of the following three determinants,
\begin{eqnarray}\label{008}
&& \tau_{n}^{(2)}=\det_{1\leq i,j \leq N}\left( m_{2i-1, 2j-2}^{(n)} \right), \\
&& \tau_{n}^{(3)}=\det_{1\leq i,j \leq N}\left( m_{2i-2, 2j-1}^{(n)} \right),  \\
&& \tau_{n}^{(4)}=\det_{1\leq i,j \leq N}\left( m_{2i-2, 2j-2}^{(n)} \right),
\end{eqnarray}
still satisfies the condition (\ref{007}). More importantly, if we combine the matrix elements of these four determinants into a $2\times 2$-block determinant
\[ \label{taublock}
\tau_{n}=\left| \begin{array}{cc} \left(m_{2i-1, 2j-1}^{(n)}\right)_{1\le i\le N_1, 1\le j\le M_1}  & \left( m_{2i-1, 2j-2}^{(n)} \right)_{1\le i\le N_1, 1\le j\le M_2} \\
\left( m_{2i-2, 2j-1}^{(n)} \right)_{1\le i\le N_2, 1\le j\le M_1} & \left( m_{2i-2, 2j-2}^{(n)} \right)_{1\le i\le N_2, 1\le j\le M_2} \end{array}\right|,
\]
where $\left[ N_{1},  N_{2}, M_{1},  M_{2}\right]$ are arbitrary non-negative integers with  $N_{1}+N_{2}=M_{1}+M_{2}$, then this $\tau_{n}$ would still satisfy the condition  (\ref{007}). This result is summarized in the following lemma, which can be regarded as a generalization of Lemma 3.2 in \cite{OhtaJY2012}. This more general determinant solution fulfilling the dimension reduction condition (\ref{007}) will open the way for us to discover wider classes of rogue waves in the nonlocal NLS equation (\ref{eq1:PTNLS}).

\textbf{Lemma 1}  \emph{Defining the matrix element $m^{(n)}_{i,j}$ as}
\[
m_{i,j}^{(n)}= \mathcal{A}_{i} \mathcal{B}_{j} m^{(n)}|_{p=1, \ q=1}, \label{lemma1-001}
\]
where
\[
m^{(n)}=\frac{1}{p+q}\left(-\frac{p}{q}\right)^{n}e^{\xi+\eta}, \  \xi=p x_{1}+p^2x_{2}, \ \eta= q x_{1}-q^2x_{2}, \label{lemma1-002}
\]
\emph{and $\mathcal{A}_{i}$, $\mathcal{B}_{j}$ are as defined in (\ref{003}), then the $2\times 2$-block determinant (\ref{taublock}), i.e., }
\[ \label{lemma1solution}
\tau_{n}=\left|
\begin{array}{cccccc}
  m_{1,\hspace{0.05cm} 1}^{(n)} & \cdots & m_{1,\hspace{0.05cm} 2 M_{1}-1}^{(n)} & m_{1,\hspace{0.05cm} 0}^{(n)} & \cdots & m_{1,\hspace{0.05cm} 2M_{2}-2}^{(n)} \\
  \vdots &    &   \vdots & \vdots &    &   \vdots  \\
  m_{2N_{1}-1,\hspace{0.05cm} 1}^{(n)} & \cdots & m_{2N_{1}-1,\hspace{0.05cm} 2M_1-1}^{(n)} & m_{2N_1-1,\hspace{0.05cm} 0}^{(n)} & \cdots & m_{2N_1-1,\hspace{0.05cm} 2M_2-2}^{(n)} \\
   m_{0,\hspace{0.05cm} 1}^{(n)} & \cdots & m_{0,\hspace{0.05cm} 2 M_{1}-1}^{(n)} & m_{0,\hspace{0.05cm} 0}^{(n)} & \cdots & m_{0,\hspace{0.05cm} 2M_{2}-2}^{(n)} \\
  \vdots &    &   \vdots & \vdots &    &   \vdots  \\
  m_{2N_{2}-2,\hspace{0.05cm} 1}^{(n)} & \cdots & m_{2N_{2}-2,\hspace{0.05cm} 2M_1-1}^{(n)} & m_{2N_2-2,\hspace{0.05cm} 0}^{(n)} & \cdots & m_{2N_2-2,\hspace{0.05cm} 2M_2-2}^{(n)}
\end{array}
\right|,
\]
\emph{with $N_{1}+N_{2}=M_{1}+M_{2}$, satisfies the $(1+1)$-dimensional bilinear equations}
\begin{eqnarray}\label{1p1blineareqs}
\left.
\begin{array}{c}
 \left(D_{x_{1}}^2 + 2\right)\tau_{n}\cdot\tau_{n}=2 \tau_{n+1} \tau_{n-1} \\
 \left(D_{x_{1}}^2-D_{x_{2}}\right)\tau_{n+1} \cdot \tau_{n}=0.
\end{array}
\right\}
\end{eqnarray}

\emph{Proof.} First of all, since the $(2+1)$-dimensional determinant (\ref{dimentionalindex}) with arbitrary indices (\ref{arbiindex}) satisfies the $(2+1)$-dimensional bilinear equations (\ref{004}) \cite{OhtaJY2012}, this determinant with a particular choice of indices as in (\ref{lemma1solution}) certainly also satisfies these bilinear equations. Thus, if this determinant also satisfies the dimension reduction condition (\ref{007}), then it would satisfy the $(1+1)$-dimensional bilinear equations (\ref{1p1blineareqs}). Meanwhile, by taking $x_{-1}=0$, this $(2+1)$-dimensional determinant reduces to the $(1+1)$-dimensional determinant (\ref{lemma1solution}).

To show the $(2+1)$-dimensional determinant (\ref{dimentionalindex}) with index choices of (\ref{lemma1solution}) satisfies the dimension reduction condition (\ref{007}), we recall that
$m_{i,j}^{(n)}$ with $p=1$ and $q=1$ and arbitrary indices $(i, j)$ satisfies the contiguity condition~\cite{OhtaJY2012}
\begin{eqnarray}\label{006}
(\partial_{x_{1}}+\partial_{x_{-1}}) \left(m_{i,j}^{(n)}|_{p=1,q=1} \right)=
2\sum^i_{k=0,\ k:even}\frac{1}{k!}m_{i-k,j}^{(n)}|_{p=1,q=1}+2\sum_{l=0,\ l:even}^{j}\frac{1}{l!}m_{i,j-l}^{(n)}|_{p=1,q=1}.
\end{eqnarray}
Then, using the cofactor expansion of determinants, we get
\begin{eqnarray}
&& (\partial_{x_{1}}+\partial_{x_{-1}}) \tau_{n} = \sum_{i=1}^{2}\sum_{j=1}^{2}
\sum_{k=1}^{N_{i}}\sum_{l=1}^{M_{j}}\Delta_{k,l}^{(i,j)}\left(\partial_{x_{1}}+\partial_{x_{-1}}\right) \left(m_{2k-i, \, 2l-j}^{(n)} \big|_{p=1,q=1}\right)   \nonumber \\
&& =2\sum_{i=1}^{2}\sum_{j=1}^{2}
\sum_{k=1}^{N_{i}}\sum_{l=1}^{M_{j}}\Delta_{k,l}^{(i,j)}\left[\sum^{2k-i}_{\alpha=0,\ \alpha:even}\frac{1}{\alpha!}m_{2(k-\frac{\alpha}{2})-i, \ 2l-j}^{(n)}\big|_{p=1,q=1}+\sum^{2l-j}_{\beta=0,\ \beta:even}\frac{1}{\beta!}m_{2k-i,\ 2(l-\frac{\beta}{2})-j}^{(n)}\big|_{p=1,q=1}\right],
\end{eqnarray}
where $\Delta_{k,l}^{(i,j)}$ is the cofactor of the element $m_{2k-i, \, 2l-j}^{(n)}$ in the determinant (\ref{lemma1solution}). In these summations, only the terms with $\alpha=0$ and $\beta=0$ survive, because the other terms correspond to determinants of the type (\ref{lemma1solution}) but with two identical rows or columns. As a result,
\begin{eqnarray}
&& (\partial_{x_{1}}+\partial_{x_{-1}}) \tau_{n} = 2\sum_{i=1}^{2}\sum_{j=1}^{2}
\sum_{k=1}^{N_{i}}\sum_{l=1}^{M_{j}}\Delta_{k,l}^{(i,j)}\left[m_{2k-i, \ 2l-j}^{(n)}\big|_{p=1,q=1}+m_{2k-i,\ 2l-j}^{(n)}\big|_{p=1,q=1}\right]=4(N_1+N_2)\tau_{n}.
\end{eqnarray}
Thus, the dimension reduction condition is satisfied, and the determinant (\ref{lemma1solution}) then satisfies the $(1+1)$-dimensional bilinear equations (\ref{1p1blineareqs}). $\blacksquare$

From Lemma 1, by taking $x_{1}=x$ and $x_{2}=-\textrm{i}t$, then
\[
f=\tau_{0}, \quad g=\tau_{1},  \quad h=\tau_{-1}
\]
would satisfy the $(1+1)$-dimensional bilinear equations (\ref{Bilineform13}). But these $(f, g, h)$ functions are not just polynomials of $x$ and $t$. Rather, they are polynomials multiplying $(m^{(n)})^{N_1+N_2}$, which is an exponential of a linear
function of $x$ and $t$. However, the bilinear equations (\ref{Bilineform13}) are invariant when $(f, g, h)$ are multiplied by an
exponential of a linear function in $(x,t)$. Thus, when we define
\[
\sigma_{n}=\frac{\tau_{n}}{(m^{(n)})^{N_1+N_2}},
\]
then
\[ \label{fgh}
f=\sigma_{0}, \quad g=\sigma_{1},  \quad h=\sigma_{-1}
\]
would be polynomials of $x$ and $t$ but still satisfy the $(1+1)$-dimensional bilinear equations (\ref{Bilineform13}).

\subsection{Nonlocal reduction condition}
Now we consider the nonlocal reduction condition (\ref{NonlocPara}). In view of Eq. (\ref{fgh}), this condition is
\[\label{NonlocRedcTaun}
\bar{\sigma}_{0}=\sigma_{0},\ \ \bar{\sigma}_{-1}=\sigma_{1}.
\]
Due to the nonlocality of this condition, i.e., the $\sigma_n$ functions at $x$ and $-x$ need to be related, imposition of this condition would require the use of explicit algebraic expressions of the $\sigma_n$ solutions, which we will derive first.

Let $\widehat{m}_{i,j}^{(n)} \equiv m_{i,j}^{(n)}/m^{(n)}$, where $m_{i,j}^{(n)}$ and $m^{(n)}$ are as given in Eqs. (\ref{lemma1-001})-(\ref{lemma1-002}) with $x_{1}=x$ and $x_{2}=-\textrm{i}t$.
Then $\widehat{m}_{i,j}^{(n)}$ can be expressed as \cite{OhtaJY2012}
\[\label{Schurpolyexps}
\widehat{m}_{i,j}^{(n)} = \sum_{\nu=0}^{\min(i,j)} \frac{1}{4^{\nu}} \sum_{k=0}^{i-\nu}\sum_{l=0}^{j-\nu}
a_{k} b_{l} S_{i-k-\nu}(\textbf{\emph{x}}^{+}(n)+\nu \textbf{\emph{s}}) S_{j-l-\nu}(\textbf{\emph{x}}^{-}(n)+\nu \textbf{\emph{s}}),
\]
where $S_{i-k-\nu}(\textbf{\emph{x}}^{+}(n)+\nu \textbf{\emph{s}})$ and $S_{j-l-\nu}(\textbf{\emph{x}}^{-}(n)+\nu \textbf{\emph{s}})$ are Schur polynomials with vectors $\textbf{\emph{x}}^{\pm}(n)$ and $\textbf{\emph{s}}$ defined in equations (\ref{skrkexpcoeff})-(\ref{skrkexpcoeff2}). Using the formula of $\tau_n$ in (\ref{lemma1solution}) and the above
algebraic representation, the expression of $\sigma_n$ in Eqs. (\ref{Sigma-n})-(\ref{phipsiexp}) will be obtained.  Moreover,  $\sigma_{n}$ can be rewritten as the following $3N\times 3N$ determinant~\cite{OhtaJY2012}
\begin{eqnarray}\label{Bigdetermin}
&& \sigma_{n}= \det_{1 \leq i, j\leq 2} \left[ mat_{
\begin{subarray}{l}
1\leq k_i \leq N_{i} \\
1\leq l_j \leq M_{j}
\end{subarray}}
\left(\sum_{\nu=0}^{\min\left(2k_i-i, \ 2l_j-j\right)} \Phi_{2k_i-i, \ \nu}^{(n)}\Psi_{2l_j-j, \ \nu}^{(n)}
\right) \right]=
\det_{1 \leq i, j\leq 2} \left[ mat_{
\begin{subarray}{l}
1\leq k_i \leq N_{i} \\
1\leq l_j \leq M_{j}
\end{subarray}}
\left(\sum_{\nu=0}^{2N-1} \Phi_{2k_i-i, \ \nu}^{(n)}\Psi_{2l_j-j, \ \nu}^{(n)}
\right) \right] \nonumber \\
&& \hspace{-1cm} =(-1)^{N}\left|
\begin{array}{cccc}
O_{N \times N} &
\begin{array}{cccc}
\Phi_{10}^{(n)} & \Phi_{11}^{(n)}  & \cdots & \Phi_{1, 2N-1}^{(n)}  \\
 \vdots &  \vdots  &   &  \vdots \\
\Phi_{2N_{1}-1, 0}^{(n)} & \Phi_{2N_{1}-1, 1}^{(n)}  & \cdots & \Phi_{2N_{1}-1, 2N-1}^{(n)}  \\
\Phi_{00}^{(n)} & \Phi_{01}^{(n)}  & \cdots & \Phi_{0, 2N-1}^{(n)}  \\
 \vdots &  \vdots  &   &  \vdots \\
\Phi_{2N_{2}-2, 0}^{(n)} & \Phi_{2N_{2}-2, 1}^{(n)}  & \cdots & \Phi_{2N_{2}-2, 2N-1}^{(n)}
\end{array} \\
 \begin{array}{cccccc}
\Psi_{10}^{(n)}  & \cdots & \Psi_{2M_{1}-1, 0}^{(n)} & \Psi_{00}^{(n)}  & \cdots & \Psi_{2M_{2}-2, 0}^{(n)} \\
\Psi_{11}^{(n)}  & \cdots & \Psi_{2M_{1}-1, 1}^{(n)} & \Psi_{01}^{(n)}  & \cdots & \Psi_{2M_{2}-2, 1}^{(n)}  \\
\vdots  &    & \vdots &   \vdots &  & \vdots  \\
\Psi_{1,2N-1}^{(n)}  & \cdots & \Psi_{2M_{1}-1, 2N-1}^{(n)} & \Psi_{0, 2N-1}^{(n)}  & \cdots & \Psi_{2M_{2}-2, 2N-1}^{(n)}
\end{array} & I_{2N \times 2N}
\end{array}
 \right|,
\end{eqnarray}
where we have defined $N\equiv N_1+N_2$ and $\Phi_{i, \nu}^{(n)}=\Psi_{i, \nu}^{(n)}\equiv 0$ when $i<\nu$, $O_{N \times N}$ is the zero matrix of size $N \times N$, and $I_{2N \times 2N}$ is the identify matrix of size $2N \times 2N$.

To impose the nonlocal reduction condition (\ref{NonlocRedcTaun}), we also need two properties on Schur polynomials, which are provided in the following two lemmas.

\textbf{Lemma 2.}   \emph{Schur polynomials $S_{k}(\textbf{x})$ with $ \textbf{x}=\left( x_{1}, x_{2}, x_{3}, \cdots \right)$ and Schur polynomials $S_{k}(\textbf{y})$ with
\[
\textbf{y}=\left( -x_{1}, x_{2}, -x_{3},\cdots\right)=\{(-1)^j x_{j}\}_{j=1}^{\infty}
\]
are related as
\[
S_{k}(\textbf{y})=(-1)^k S_{k}(\textbf{x}).
\]
}

\textbf{Lemma 3.}  \emph{Schur polynomials $S_{k}(\hat{\textbf{x}})$ with $\hat{\textbf{x}}=\left( x_{1}, x_{2}, 0, x_{4}, 0, x_6, 0, \cdots  \right)$ can be expressed as a series of Schur polynomials $S_{k}(\hat{\textbf{y}})$ with $\hat{\textbf{y}}=\left(x_{2}, x_{4}, x_6,\cdots \right)$ as
\[
S_{k}(\hat{\textbf{x}})=\sum_{l=0}^{\left[ \frac{k}{2} \right]}\frac{(x_{1})^{k-2l}}{(k-2l)!}S_{l}(\hat{\textbf{y}}),
\]
where $\left[ x \right]$ represents the integer part of a real number $x$. }

Both lemmas can be proved through a direct calculation using the definition (\ref{def_Schur}) of Schur polynomials.

With the above algebraic expressions of $\sigma_{n}$ and the two lemmas, we now derive parameter constraints on $a_{k}$  and $b_{l}$ so that $\sigma_{n}$ satisfies the nonlocal reduction condition (\ref{NonlocRedcTaun}), or equivalently
\[ \label{NonlocsigReduc}
\sigma_{n}= \overline{\sigma}_{-n}.
\]

First, if we impose nonlocal reductions on the vector $\textbf{\emph{x}}^{\pm}(n)$ as defined in (\ref{skrkexpcoeff}), we get
\begin{eqnarray}
&& \overline{x}_{1}^{\pm}(-n)= -x_{1}^{\pm}(n)-1; \quad \overline{x}_{k}^{\pm}=-x_{k}^{\pm}-2r_{k}, \ (k > 1).
\end{eqnarray}
Thus,
\[
\overline{\textbf{\emph{x}}}^{\pm}(-n)+\nu\ \textbf{\emph{s}}=\textbf{y}^{\pm}(n)+\nu\ \textbf{\emph{s}}+\hat{\textbf{y}}^{\pm},
\]
where vectors $\textbf{y}^{\pm}$ and $\hat{\textbf{y}}^{\pm}$ are defined as
\[
\textbf{y}^{\pm}(n)= \left(-x_{1}^{\pm}(n), x_{2}^{\pm}, -x_{3}^{\pm}, x_{4}^{\pm}, \cdots\right), \quad
\hat{\textbf{y}}^{\pm}=\left(-1, -2x_{2}^{\pm}-2r_{2}, 0, -2x_{4}^{\pm}-2r_{4}, 0,  \cdots\right).
\]
Here, the fact of $r_1=r_3=\cdots=r_{odd}=0$ has been used~\cite{OhtaJY2012}. Since
\begin{eqnarray} \label{NonlocSchurPoly}
&& \sum_{k=0}^{\infty} S_{k}\left(\overline{\textbf{\emph{x}}}^{\pm}(-n)+\nu\ \textbf{\emph{s}}\right)\lambda^{k} =\sum_{k=0}^{\infty} S_{k}\left(\textbf{y}^{\pm}(n)+\nu\ \textbf{\emph{s}}+\hat{\textbf{y}}^{\pm}\right)\lambda^{k},  \nonumber \\
&& = \exp\left(\sum_{j=1}^{\infty} \left(y_j^{\pm}(n)+\nu\ s_j+\hat{y}_j^{\pm}\right)\lambda^j\right)
=\exp\left(\sum_{j=1}^{\infty} \left(y_j^{\pm}(n)+\nu\ s_j\right)\lambda^j\right)\exp\left(\sum_{j=1}^{\infty} \hat{y}_j^{\pm}\lambda^j\right) \nonumber \\
&& = \sum_{k=0}^{\infty} S_{k}(\textbf{y}^{\pm}(n)+\nu \textbf{\emph{s}})\lambda^{k} \sum_{k=0}^{\infty} S_{k}(\hat{\textbf{y}}^{\pm}) \lambda^{k}=\sum_{k=0}^{\infty} \sum_{\mu_{1}+\mu_{2}=k} S_{\mu_{1}}(\textbf{y}^{\pm}(n)+\nu \textbf{\emph{s}})S_{\mu_{2}}(\hat{\textbf{y}}^{\pm}) \lambda^{k},
\end{eqnarray}
by using the above two lemmas and the fact of $s_1=s_3=\cdots=s_{odd}=0$ \cite{OhtaJY2012}, comparison of the power of $\lambda^{k}$ on the two sides of equation (\ref{NonlocSchurPoly}) gives
\begin{eqnarray}\label{comparpowern}
S_{k}\left(\overline{\textbf{\emph{x}}}^{\pm}(-n)+\nu\ \textbf{\emph{s}}\right)= (-1)^{k} \sum_{\mu=0}^{k} S_{\mu}(\textbf{\emph{x}}^{\pm}(n)+\nu \textbf{\emph{s}}) \sum_{l=0}^{\left[ \frac{k-\mu}{2} \right]}\frac{S_{l}(\textbf{\emph{w}})}{(k-\mu-2l)!},
\end{eqnarray}
where $\textbf{\emph{w}}=\left( -2x_{2}^{\pm}-2r_{2}, -2x_{4}^{\pm}-2r_{4},\cdots \right)$. Then by switching the order of summations, the above equation becomes
\begin{eqnarray}\label{Phatpolynomial}
S_{k}\left(\overline{\textbf{\emph{x}}}^{\pm}(-n)+\nu\ \textbf{\emph{s}}\right)=(-1)^{k} \sum_{l=0}^{\left[ \frac{k}{2} \right]} S_{l}(\textbf{\emph{w}}) \sum_{\mu=0}^{k-2l} \frac{S_{\mu}(\textbf{\emph{x}}^{\pm}(n)+\nu \textbf{\emph{s}}) }{(k-2l-\mu)!}.
\end{eqnarray}
Using this formula and imposing nonlocal reductions on functions $\Phi_{i,\nu}^{(n)}$ and $\Psi_{i,\nu}^{(n)}$ as defined in Eq. (\ref{phipsiexp}), we get
\begin{eqnarray}
&& \overline{\Phi_{i,\nu}^{(-n)}}=\frac{1}{2^{\nu}}\sum_{k=0}^{i-\nu}a_{k}^*S_{i-\nu-k}\left(\overline{\textbf{\emph{x}}^{+}}(-n)+\nu\ \textbf{\emph{s}}\right) \nonumber \\
&&=\frac{1}{2^{\nu}} \sum_{k=0}^{i-\nu} a_{k}^* (-1)^{i-\nu-k} \sum_{l=0}^{\left[ \frac{i-\nu-k}{2} \right]} S_{l}(\textbf{\emph{w}})  \sum_{\mu=0}^{i-\nu-k-2l} \frac{S_{\mu}(\textbf{\emph{x}}^{+}(n)+\nu \textbf{\emph{s}}) }{(i-\nu-k-2l-\mu)!}.  \nonumber
\end{eqnarray}
Switching the order of summations then leads to
\[ \label{Phifunctions}
\overline{\Phi_{i,\nu}^{(-n)}}=\frac{(-1)^{i-\nu}}{2^{\nu}} \sum_{l=0}^{\left[\frac{i-\nu}{2}\right]} S_{l}(\textbf{\emph{w}})   \left[\sum_{k=0}^{i-\nu-2l} (-1)^{k} a_{k}^* \sum_{\mu=0}^{i-\nu-2l-k} \frac{S_{\mu}(\textbf{\emph{x}}^{+}(n)+\nu \textbf{\emph{s}}) }{(i-\nu-2l-k-\mu)!}\right].
\]
Similarly, we get
\[\label{Psifunctions}
\overline{\Psi_{j,\nu}^{(-n)}}=
\frac{(-1)^{j-\nu}}{2^{\nu}} \sum_{l=0}^{\left[\frac{j-\nu}{2}\right]} S_{l}(\textbf{\emph{w}})    \left[\sum_{k=0}^{j-\nu-2l} (-1)^{k} b_{k}^* \sum_{\mu=0}^{j-\nu-2l-k} \frac{S_{\mu}(\textbf{\emph{x}}^{-}(n)+\nu \textbf{\emph{s}}) }{(j-\nu-2l-k-\mu)!}\right].
\]
These two expressions can be reshaped into matrix forms:
\begin{eqnarray}
&& \overline{\Phi_{i,\nu}^{(-n)}}=\frac{(-1)^{i-\nu}}{2^{\nu}} \sum_{l=0}^{\left[\frac{i-\nu}{2}\right]} S_{l}(\textbf{\emph{w}})
\left[
\begin{array}{c}S_{i-\nu-2l}(\textbf{\emph{z}}^+),
S_{i-\nu-2l-1}(\textbf{\emph{z}}^+),
\cdots, S_{0}(\textbf{\emph{z}}^+)
\end{array} \right]
M_{i-\nu-2l}
\left(
  \begin{array}{c}
    a_{0}^* \\
    a_{1}^* \\
    \vdots \\
    a_{i-\nu-2l}^* \\
  \end{array}
\right), \label{linearmatrixform1} \\
&& \overline{\Psi_{j,\nu}^{(-n)}}=\frac{(-1)^{j-\nu}}{2^{\nu}} \sum_{l=0}^{\left[\frac{j-\nu}{2}\right]} S_{l}(\textbf{\emph{w}})
\left[
\begin{array}{c}
 S_{j-\nu-2l}(\textbf{\emph{z}}^{-}), S_{j-\nu-2l-1}(\textbf{\emph{z}}^{-}), \cdots,  S_{0}(\textbf{\emph{z}}^{-})
\end{array} \right]
M_{j-\nu-2l}
\left(
  \begin{array}{c}
    b_{0}^* \\
    b_{1}^* \\
    \vdots \\
    b_{j-\nu-2l}^* \\
  \end{array}
\right),  \label{linearmatrixform2}
\end{eqnarray}
where $\textbf{\emph{z}}^\pm\equiv \textbf{\emph{x}}^{\pm}(n)+\nu \textbf{\emph{s}}$, and the square matrices $M_{i-\nu-2l}$, $M_{j-\nu-2l}$ are defined as
\begin{eqnarray}
&&M_{j}=\left(
  \begin{array}{cccc}
    1/0! & 0 & \cdots & 0 \\
    1/1! & -1/0! & \cdots & 0 \\
     \vdots & \vdots & \ddots & \vdots \\
    1/j! & -1/(j-1)! & \cdots & (-1)^{j}/0! \\
  \end{array}
\right).
\end{eqnarray}
In view of the lower triangular structure of the matrix $M_j$, if parameters $\{a_j, b_j\}$ satisfy the conditions
\begin{eqnarray}\label{ParaCondition1}
M_{j}
\left(
  \begin{array}{c}
    a_{0}^*\\ a_{1}^*\\ a_{2}^* \\ a_{3}^* \\ \vdots\\ a_{j}^*
    \end{array}
\right)= \alpha_0 \left(
  \begin{array}{c}
    a_{0} \\    a_{1} \\ a_2 \\ a_3 \\
    \vdots \\
    a_{j} \end{array}
\right)+
\alpha_1 \left(
  \begin{array}{c}
    0 \\
    0 \\ a_0 \\ a_1 \\
    \vdots \\
    a_{j-2} \end{array}
\right)+\cdots + \alpha_{\left[\frac{j}{2}\right]} \left(
  \begin{array}{c}
    0 \\
    0 \\ 0 \\ 0 \\
    \vdots \\
    a_{j-2\left[\frac{j}{2}\right]} \end{array}
\right),
\end{eqnarray}
and
\begin{eqnarray}\label{ParaCondition2}
M_{j}
\left(
  \begin{array}{c}
    b_{0}^*\\ b_{1}^*\\ b_{2}^* \\ b_{3}^* \\ \vdots\\ b_{j}^*
    \end{array}
\right)=\beta_0 \left(
  \begin{array}{c}
    b_{0} \\    b_{1} \\ b_2 \\ b_3 \\
    \vdots \\
    b_{j} \end{array}
\right)+
\beta_1 \left(
  \begin{array}{c}
    0 \\
    0 \\ b_0 \\ b_1 \\
    \vdots \\
    b_{j-2} \end{array}
\right)+\cdots + \beta_{\left[\frac{j}{2}\right]} \left(
  \begin{array}{c}
    0 \\
    0 \\ 0 \\ 0 \\
    \vdots \\
    b_{j-2\left[\frac{j}{2}\right]} \end{array}
\right),
\end{eqnarray}
where $\alpha_0=\beta_0=1$, and $\alpha_1, \alpha_2, \dots$, $\beta_1, \beta_2, \dots$ are some constants, then the right-hand sides of (\ref{linearmatrixform1})-(\ref{linearmatrixform2}) would become
\begin{eqnarray}
\label{linearconnnction1}
&&\overline{\Phi_{i,\nu}^{(-n)}}=(-1)^{i-\nu} \sum_{l=0}^{\left[\frac{i-\nu}{2}\right]} \left[\sum_{k=0}^{l} \alpha_k S_{l-k}(\textbf{\emph{w}})\right] \Phi_{i-2l,\nu}^{(n)},\\
 \label{linearconnnction2}
&&\overline{\Psi_{j,\nu}^{(-n)}}=(-1)^{j-\nu} \sum_{l=0}^{\left[\frac{j-\nu}{2}\right]} \left[\sum_{k=0}^{l} \beta_k S_{l-k}(\textbf{\emph{w}})\right] \Psi_{j-2l,\nu}^{(n)}.
\end{eqnarray}
These two equations show that, for any indices $(i, \nu)$ and $(j, \nu)$, the nonlocal reduction function $\overline{\Phi_{i,\nu}^{(-n)}}$ can be expressed as a linear combination of functions $\Phi_{i-2l,\nu}^{(n)}$ for $l=0,1,\cdots$, and the nonlocal reduction function $\overline{\Psi_{j,\nu}^{(-n)}}$ can be expressed as a linear combination of functions $\Psi_{j-2l,\nu}^{(n)}$ for $l=0,1,\cdots$. Then,  using these relations and performing simple determinant manipulations on the $3N\times 3N$ determinant (\ref{Bigdetermin}), one can quickly show that
\[
\overline{\sigma}_{-n} = (-1)^{N_{1}+M_{2}} \sigma_{n}.
\]
Thus, by redefining $(-1)^{(N_{1}+M_{2})/2} \sigma_{n}$ as a new $\sigma_{n}$ function, the nonlocal reduction condition (\ref{NonlocsigReduc}) would be satisfied.

Our last task is to solve the equations (\ref{ParaCondition1})-(\ref{ParaCondition2}) and derive explicit conditions on the parameters $\{a_j, b_j\}$ so that the nonlocal reduction condition (\ref{NonlocsigReduc}) would be satisfied.

To proceed, we notice that by performing simple manipulations to the $3N\times 3N$ determinant (\ref{Bigdetermin}) of $\sigma_n$ as was done in \cite{OhtaJY2012}, we can set
\[
a_0=b_0=1, \quad a_2=a_4=\cdots=a_{even}=0, \quad b_2=b_4=\cdots=b_{even}=0
\]
without any loss of generality. Under this parameter normalization, by setting $j=2k-1$ in conditions (\ref{ParaCondition1}), these conditions on parameters $\{a_1, a_3, \dots\}$ can be written simply as
\begin{eqnarray}
&& \frac{1}{(2k-2)!}-\frac{1}{(2k-3)!}a_1^*-\frac{1}{(2k-5)!}a_3^*-\cdots-\frac{1}{1!}a_{2k-3}^*=\alpha_{k-1},   \\
&& \frac{1}{(2k-1)!}-\frac{1}{(2k-2)!}a_1^*-\frac{1}{(2k-4)!}a_3^*-\cdots-\frac{1}{0!}a_{2k-1}^*=\alpha_0 a_{2k-1}+\alpha_1a_{2k-3}+\cdots+\alpha_{k-1}a_1,    \end{eqnarray}
where $k=1, 2, \dots$. Substituting the first equation into the second and rearranging terms, we then obtain the recurrence relation (\ref{a_recur}) for $\Re(a_{2k-1})$ in Theorem 1, and $\Im(a_{2k-1})$ are free parameters. Notice that the double summation term in (\ref{a_recur}) is always real, because its complex conjugate can be shown to be equal to itself after switching the two summations. Performing similar calculations, we can derive the recurrence relation (\ref{b_recur}) for $\Re(b_{2k-1})$ in Theorem 1, and $\Im(b_{2k-1})$ are free parameters. This completes the proof of Theorem 1 (except for the boundary condition part).

\subsection{Boundary conditions and proof of Theorem 2}
In order to prove the boundary conditions (\ref{BoundaryCond1}) in Theorem 1 and the results in Theorem 2, we need to derive the highest-power terms for the polynomials $\sigma_0$ and $\sigma_1$ in the solution (\ref{Th1-solutions}). This calculation follows the same approach as in \cite{OhtaJY2012}. But since the determinant (\ref{Sigma-n}) of the present $\sigma_n$ function has a different structure, results of this calculation will be quite different.

We start by applying the Laplace expansion to the $3N \times 3N$ determinant (\ref{Bigdetermin}) of $\sigma_{n}$ and get
\begin{eqnarray}
&&\sigma_{n}=\sum_{0\leq \nu_{1} < \nu_{2}<\cdots < \nu_{N}\leq 2N-1} \left| \begin{array}{cccc}
\Phi_{1,\nu_{1}}^{(n)} & \Phi_{1, \nu_{2}}^{(n)}  & \cdots & \Phi_{1, \nu_{N}}^{(n)}  \\
 \vdots &  \vdots  &   &  \vdots \\
\Phi_{2N_{1}-1, \nu_{1}}^{(n)} & \Phi_{2N_{1}-1, \nu_{2}}^{(n)}  & \cdots & \Phi_{2N_{1}-1, \nu_{N}}^{(n)}  \\
\Phi_{0, \nu_{1}}^{(n)} & \Phi_{0, \nu_{2}}^{(n)}  & \cdots & \Phi_{0, \nu_{N}}^{(n)}  \\
 \vdots &  \vdots  &   &  \vdots \\
\Phi_{2N_{2}-2, \nu_{1}}^{(n)} & \Phi_{2N_{2}-2, \nu_{2}}^{(n)}  & \cdots & \Phi_{2N_{2}-2, \nu_{N}}^{(n)}
\end{array}\right| \nonumber \\
&& \hspace{0.5cm} \times
\left| \begin{array}{cccccc}
\Psi_{1,\nu_{1}}^{(n)}  & \cdots & \Psi_{2M_{1}-1,\nu_{1}}^{(n)} & \Psi_{0,\nu_{1}}^{(n)}  & \cdots & \Psi_{2M_{2}-2,\nu_{1}}^{(n)} \\
\Psi_{1,\nu_{2}}^{(n)}  & \cdots & \Psi_{2M_{1}-1, \nu_{2}}^{(n)} & \Psi_{0,\nu_{2}}^{(n)}  & \cdots & \Psi_{2M_{2}-2, \nu_{2}}^{(n)}  \\
\vdots  &    & \vdots &   \vdots &  & \vdots  \\
\Psi_{1,\nu_{N}}^{(n)}  & \cdots & \Psi_{2M_{1}-1, \nu_{N}}^{(n)} & \Psi_{0,\nu_{N}}^{(n)}  & \cdots & \Psi_{2M_{2}-2, \nu_{N}}^{(n)}
\end{array}
\right|. \label{LapaceExpdet}
\end{eqnarray}
Since the degrees of polynomial functions $\Phi_{i, \nu}^{(n)}$ and $\Psi_{j, \nu}^{(n)}$ are equal to $i-\nu$ and $j-\nu$, with the highest-power terms being $a_{0}(x-2\textrm{i}t)^{i-\nu}/\left[(i-\nu)! 2^\nu\right]$ and $b_{0}(x+2\textrm{i}t)^{j-\nu}/\left[(j-\nu)! 2^\nu\right]$ \cite{OhtaJY2012}, the highest-power terms in the above equation will come from the choice of
$\nu_{1}=0$, $\nu_{2}=1$, $\ldots$, $\nu_{N}=N-1$ in the summation. Under this choice, replacing every $\Phi_{i, \nu}^{(n)}$ and $\Psi_{j, \nu}^{(n)}$ by their highest-power terms and rearranging rows of the $|\Phi|$ determinant and columns of the $|\Psi|$ determinant, we can derive the highest-power terms in the $\sigma_n$ polynomial function. Details of this calculation depend on the values of $[N_1, N_2]$ and $[M_1, M_2]$. Let us consider the case of $N_{1}> N_{2}$ as an example. In this case, the $|\Phi|$ determinant in Eq. (\ref{LapaceExpdet}) under the above choice of $\nu_k=k-1$ becomes
\begin{equation}
\det \left(\Phi_{i,j}^{(n)}\right)=(-1)^{N_{2}\left(2N_{1}-N_{2}-1\right)/2}
\left|
\begin{array}{cc}
  P_1 & O\\
  P_2 & P_3
\end{array}
\right| + (\text{lower degree terms}),
\end{equation}
where $O$ is the zero matrix of size $(2N_2) \times (N_1-N_2)$,
\begin{eqnarray}
&&P_1=\left[
\begin{array}{ccccc}
 a_{0}  & 0 & 0 & \cdots & 0 \\
  a_{0} x_{1}^+ & \frac{a_{0}}{2} & 0 & \cdots & 0 \\
  \frac{a_{0} (x_{1}^+)^2}{2!} & \frac{a_{0}x_{1}^+}{1! 2}& \frac{a_{0}}{2^2} & \cdots & 0  \\
  \vdots & \vdots & \vdots & \ddots &\vdots \\
  \frac{a_{0}(x_{1}^{+})^{2N_{2}-1}}{(2N_{2}-1)!} & \frac{a_{0}(x_{1}^{+})^{2N_{2}-2}}{(2N_{2}-2)!2} & \frac{a_{0}(x_{1}^{+})^{2N_{2}-3}}{(2N_{2}-3)!2^2} & \cdots & \frac{a_{0}}{2^{2N_{2}-1}} \end{array}
\right],
\end{eqnarray}
and
\begin{eqnarray}
P_3=\left[\begin{array}{cccccc}
\frac{a_0x^{+}_1}{1!2^{2N_{2}}} &\frac{a_0}{2^{2N_{2}+1}} &0 & 0 & 0&\cdots \\
\frac{a_0(x^{+}_1)^3}{3!2^{2N_{2}}} &\frac{a_0(x^{+}_1)^2}{2!2^{2N_{2}+1}} &\frac{a_0x^{+}_1}{1!2^{2N_{2}+2}}
&\frac{a_0}{2^{2N_{2}+3}} &0  &\cdots \\
\vdots &\vdots &\vdots &\vdots & \vdots & \vdots  \\
\frac{a_0(x^{+}_1)^{2(N_1-N_2)-1}}{[2(N_1-N_2)-1]!2^{2N_{2}}} &\frac{a_0(x^{+}_1)^{2(N_1-N_2)-2}}{[2(N_1-N_2)-2]!2^{2N_{2}+1}}
&\frac{a_0(x^{+}_1)^{2(N_1-N_2)-3}}{[2(N_1-N_2)-3]!2^{2N_{2}+2}} &\frac{a_0(x^{+}_1)^{2(N_1-N_2)-4}}{[2(N_1-N_2)-4]!2^{2N_{2}+3}}
&\cdots &\frac{a_{0}(x_{1}^{+})^{N_{1}-N_{2}}}{(N_{1}-N_{2})!2^{N-1}}\end{array}\right].
\end{eqnarray}
The matrix $P_1$ is a lower triangular matrix whose determinant is a constant. The matrix $P_3$ has the same structure as in \cite{OhtaJY2012}, and thus its determinant can be calculated by the same technique as used in \cite{OhtaJY2012}. These calculations directly lead to
\begin{equation}
\det \left(\Phi_{i,j}^{(n)}\right)=\hat{c}_0 (x-2\textrm{i}t)^{(N_{1}-N_{2})(N_{1}-N_{2}+1)/2}
 + (\text{lower degree terms}),
\end{equation}
where $\hat{c}_0$ is a $(a_0, N_1, N_2)$-dependent but $n$-independent constant. When $N_{1}\le N_{2}$, $\det \left(\Phi_{i,j}^{(n)}\right)$ can be calculated similarly. In addition, $\det \left(\Psi_{i,j}^{(n)}\right)$ in Eq. (\ref{LapaceExpdet}) can be treated in a similar fashion.
Collecting these results and setting $a_0=b_0=1$ (as in Theorem 1), we find that
\begin{eqnarray}
\sigma_{n}=c_{0}(x-2\textrm{i}t)^{(N_{1}-N_{2})(N_{1}-N_{2}+1)/2}(x+2\textrm{i}t)^{(M_{1}-M_{2})(M_{1}-M_{2}+1)/2}+ (\text{lower degree terms}),
\end{eqnarray}
where $c_0$ is a $\left[N_{1}, N_{2}, M_{1}, M_{2}\right]$-dependent but $n$-independent constant.
Hence the solution (\ref{Th1-solutions}) satisfies the boundary condition (\ref{BoundaryCond1}), and Theorem 2 is also proved.

\section{Conclusions and Discussions}
In this article, we have derived wider classes of rogue wave solutions in the nonlocal $\cal{PT}$-symmetric NLS equation (\ref{eq1:PTNLS}) through the bilinear KP-reduction method, and these solutions are given explicitly as Gram determinants with matrix elements in terms of Schur polynomials. These wider classes of solutions were found through a generalization of the previous KP-reduction method, where a richer $\tau$ function structure was discovered. New rogue waves in these solutions contain not only the ones with novel polynomial degrees, but also the ones with old polynomial degrees but different functional forms. These new rogue waves were shown to exhibit distinctive solution patterns which have not been seen before.

Given these wider classes of rogue waves than those reported before \cite{YangYang2018}, a natural question is whether the nonlocal NLS equation (\ref{eq1:PTNLS}) admits even more rogue waves which are not covered in Theorem 1. This is a challenging question which is not easy to answer. Our opinion is that, in order to satisfy the dimension reduction condition (\ref{007}), our $2\times 2$-block determinant
(\ref{lemma1solution}) is the most general $\tau$ function which leads to algebraic solutions of the $(1+1)$-dimensional bilinear equations (\ref{1p1blineareqs}). Based on this opinion, we conjecture that the rogue waves reported in Theorem 1 are \emph{all} rogue-wave solutions in the $\cal{PT}$-symmetric NLS equation (\ref{eq1:PTNLS}).

Another natural question inspired by results in this article is whether the local NLS equation can also admit rogue waves with polynomial degrees beyond the type $N(N+1)$ \cite{ACA2010,OhtaJY2012}. We have examined this question and our answer is no. It is true that the more general $2\times 2$-block determinant (\ref{lemma1solution}) can also satisfy the $(1+1)$-dimensional bilinear equations of the local NLS equation before complex conjugacy reduction [see Eq. (3.6) of \cite{OhtaJY2012}, which is the counterpart of Eq. (\ref{1p1blineareqs}) in this article before the nonlocal reduction (\ref{NonlocsigReduc})]. However, to further satisfy the complex conjugacy reduction [see Eq. (3.5) of \cite{OhtaJY2012}], which is $\sigma_{n}=\sigma^*_{-n}$ in the notation of this article, we generally have to require $N_1=M_1$ and $N_2=M_2$, so that the parameter constraint of $a_k=b_k^*$ would fulfill this complex conjugacy condition. But the third example in Remark 5 shows that the solutions with $[N_1, N_2, N_1, N_2]$ are equivalent to those with $[N_1-N_2, 0, N_1-N_2, 0]$ when $N_1>N_2$ and equivalent to those with $[0, N_2-N_1, 0, N_2-N_1]$ when $N_1<N_2$. In the former case, solutions with $[N_1-N_2, 0, N_1-N_2, 0]$ are exactly the rogue waves reported in \cite{OhtaJY2012} with $N=N_1-N_2$ (see Remark 2). In the latter case, solutions with $[0, N_2-N_1, 0, N_2-N_1]$ are equivalent to those with $[N_2-N_1-1, 0, N_2-N_1-1, 0]$ in view of the second example in Remark 5, and are thus also the rogue waves reported in \cite{OhtaJY2012} with $N=N_2-N_1-1$. Thus, we conclude that the more general $2\times 2$-block determinant (\ref{lemma1solution}) does not generate new rogue waves in the local NLS equation due to the complex conjugacy reduction.

\section*{Acknowledgement}
This material is based upon work supported by the Air Force Office of Scientific
Research under award number FA9550-18-1-0098 and the National Science Foundation under award number DMS-1616122.

\begin{center}
\textbf{Appendix A: Number of free real parameters in rogue waves of Theorem 1}
\end{center}

In this appendix,  we prove the number of irreducible free real parameters given in Table 1 for rogue waves of Theorem 1. To illustrate the idea behind our proof, we use an example of solutions with index values $\left[ N_{1},  N_{2}, M_{1},  M_{2}\right]=\left[2, 1, 1, 2\right]$. Solutions with these index values correspond to the following $3\times 3$ Gram-determinant in view of Eq. (\ref{Sigma-n}),
\[\label{3b3example}
\tau_{n}=
\left|
  \begin{array}{ccc}
    m_{11}^{(n)} & m_{10}^{(n)} &  m_{12}^{(n)} \\
    m_{31}^{(n)} & m_{30}^{(n)} &  m_{32}^{(n)} \\
    m_{01}^{(n)} & m_{00}^{(n)} &  m_{02}^{(n)}
  \end{array}
\right|.
\]
Parameters contained in this determinant include $\Im(a_{1})$, $\Im(a_{3})$ and $\Im(b_{1})$ (see Theorem 1). However, we can rewrite this $\tau_{n}$ into a $9\times 9$ determinant in view of Eq. (\ref{Bigdetermin}). Then, reorganizing the rows and columns of this $9\times 9$ determinant, we get
\[\label{9b9example}
\tau_{n}=\left|
  \begin{array}{ccc}
   O_{3\times 3} & \mathcal{G}_{n} \\
   \mathcal{H}_{n} &  I_{6 \times 6}
  \end{array}
\right|,
\]
where $\mathcal{G}_{n}$ is a $3 \times 6$ matrix of functions $\Phi_{i,j}^{(n)}$,
\[
\mathcal{G}_{n}=
\left(
\begin{array}{cccccc}
  \Phi_{0,0}^{(n)} & 0 & 0& 0& 0&0 \\
  \Phi_{1,0}^{(n)} & \Phi_{1,1}^{(n)} & 0& 0& 0&0 \\
   \Phi_{3,0}^{(n)} & \Phi_{3,1}^{(n)} & \Phi_{3,2}^{(n)}& \Phi_{3,3}^{(n)}& 0&0
  \end{array}
\right),
\]
and $\mathcal{H}_{n}$ is a $6 \times 3$ matrix of functions $\Psi_{k,l}^{(n)}$,
\[
\mathcal{H}_{n}=
\left(
\begin{array}{cccccc}
  \Psi_{0,0}^{(n)} & 0 & 0& 0& 0&0 \\
  \Psi_{1,0}^{(n)} & \Psi_{1,1}^{(n)} & 0& 0& 0&0 \\
   \Psi_{2,0}^{(n)} & \Psi_{2,1}^{(n)} & \Psi_{2,2}^{(n)}& 0 & 0&0
  \end{array}
\right)^T.
\]
From the definitions of $\Phi_{k,\nu}^{(n)}$ and $\Psi_{k,\nu}^{(n)}$ in Theorem 1, we have $\Phi_{k,k}^{(n)}=\Psi_{k,k}^{(n)}=1/2^k$. Thus, performing simple determinant expansions to (\ref{9b9example}), we get
\[
\tau_{n}=\gamma\left|
\begin{array}{cccc}
  \Phi_{3,2}^{(n)} &  \Phi_{3,3}^{(n)} & 0& 0 \\
  0  & 1 & 0 & 0 \\
   0 & 0 & 1 & 0\\
   0 & 0 & 0 & 1
  \end{array}
\right|=  \gamma\ \Phi_{3,2}^{(n)},
\]
where $\gamma=\Phi_{0,0}^{(n)} \Phi_{1,1}^{(n)} \Psi_{0,0}^{(n)} \Psi_{1,1}^{(n)} \Psi_{2,2}^{(n)} = 1/16$. Notice that in this $\tau_n$ expression, parameters $\Im(a_{3})$ and $\Im(b_{1})$ have dropped out, and $\Im(a_{1})$ is the only remaining parameter in the solution (\ref{3b3example}). Following the idea of this example, we can show in general that when $N_{2}<N_{1}$, parameters $\Im(a_{2(N_{1}-N_{2})+1})$, $\Im(a_{2(N_{1}-N_{2})+3})$, $\cdots$ drop out; when $N_{2}=N_{1}$ or $N_{2}=N_{1}+1$, all $\Im(a_1)$, $\Im(a_3)$, $\cdots$ drop out; and when $N_2>N_1+1$, parameters $\Im(a_{2(N_{2}-N_{1})-1})$, $\Im(a_{2(N_{2}-N_{1})+1})$, $\cdots$ drop out. Likewise, when $M_{2}<M_{1}$, parameters $\Im(b_{2(M_{1}-M_{2})+1})$, $\Im(b_{2(M_{1}-M_{2})+3})$, $\cdots$ drop out; when $M_{2}=M_{1}$ or $M_{2}=M_{1}+1$, all $\Im(b_1)$, $\Im(b_3)$, $\cdots$ drop out; and when $M_2>M_1+1$, parameters $\Im(b_{2(M_{2}-M_{1})-1})$, $\Im(b_{2(M_{2}-M_{1})+1})$, $\cdots$ drop out. Since the nonlocal NLS equation (\ref{eq1:PTNLS}) is time-translation invariant, by a shift of the $t$ axis, we can remove one more real parameter (see Ref.~\cite{OhtaJY2012} for details). Thus, the number of irreducible free real parameters in rogue waves of Theorem 1 can be obtained as those given in Table 1.

\begin{center}
\textbf{Appendix B: Proof of Remark 5}
\end{center}

In this appendix, we prove the equivalency of solutions in Remark 5.

First, we prove that solutions with $\left[ N_{1},  N_{2}, M_{1},  M_{2}\right]= \left[\widetilde{N}_{1}, \widetilde{N}_{2}, \widetilde{M}_{1},\widetilde{M}_{2}\right]$ and those with $\left[\widetilde{N}_{2}, \widetilde{N}_{1}+1, \widetilde{M}_{2}, \widetilde{M}_{1}+1\right]$ are equivalent to each other. To prove this fact, we start from the $3N\times 3N$ determinant expression for $\sigma_{n}$ with $\left[ N_{1},  N_{2}, M_{1},  M_{2}\right]= \left[\widetilde{N}_{1}, \widetilde{N}_{2}, \widetilde{M}_{1},\widetilde{M}_{2}\right]$, which in view of Eq. (\ref{Bigdetermin}) is
\begin{eqnarray}
&&\hspace{-1.0cm}\sigma_{n}=\nonumber(-1)^{N} \times \\
&&\hspace{-1.0cm}\left|
\begin{array}{cccc}
O_{N \times N} &
\begin{array}{cccc}
\Phi_{10}^{(n)} & \Phi_{11}^{(n)}  & \cdots & \Phi_{1, 2N-1}^{(n)}  \\
 \vdots &  \vdots  &   &  \vdots \\
\Phi_{2\widetilde{N}_{1}-1, 0}^{(n)} & \Phi_{2\widetilde{N}_{1}-1, 1}^{(n)}  & \cdots & \Phi_{2\widetilde{N}_{1}-1, 2N-1}^{(n)}  \\
\Phi_{00}^{(n)} & \Phi_{01}^{(n)}  & \cdots & \Phi_{0, 2N-1}^{(n)}  \\
 \vdots &  \vdots  &   &  \vdots \\
\Phi_{2\widetilde{N}_{2}-2, 0}^{(n)} & \Phi_{2\widetilde{N}_{2}-2, 1}^{(n)}  & \cdots & \Phi_{2\widetilde{N}_{2}-2, 2N-1}^{(n)}
\end{array} \\
 \begin{array}{cccccc}
\Psi_{10}^{(n)}  & \cdots & \Psi_{2\widetilde{M}_{1}-1, 0}^{(n)} & \Psi_{00}^{(n)}  & \cdots & \Psi_{2\widetilde{M}_{2}-2, 0}^{(n)} \\
\Psi_{11}^{(n)}  & \cdots & \Psi_{2\widetilde{M}_{1}-1, 1}^{(n)} & \Psi_{01}^{(n)}  & \cdots & \Psi_{2\widetilde{M}_{2}-2, 1}^{(n)}  \\
\vdots  &    & \vdots &   \vdots &  & \vdots  \\
\Psi_{1,2N-1}^{(n)}  & \cdots & \Psi_{2\widetilde{M}_{1}-1, 2N-1}^{(n)} & \Psi_{0, 2N-1}^{(n)}  & \cdots & \Psi_{2\widetilde{M}_{2}-2, 2N-1}^{(n)}
\end{array} & I_{2N \times 2N}
\end{array}
 \right|. \label{First3Nby3N}
\end{eqnarray}
Here, $N=\widetilde{N}_{1}+\widetilde{N}_{2}=\widetilde{M}_{1}+\widetilde{M}_{2}$.

Then, we consider the similar determinant expression for
\[
\left[ N_{1},  N_{2}, M_{1},  M_{2}\right]= \left[\widetilde{N}_{2}, \widetilde{N}_{1}+1, \widetilde{M}_{2}, \widetilde{M}_{1}+1\right].
\]
Recalling that $\Phi_{0,0}^{(n)}=\Psi_{0,0}^{(n)}=1$ and $\Phi_{0,\nu}^{(n)}=\Psi_{0,\nu}^{(n)}=0$ for $\nu\ge 1$, the row of $\Phi_{0,0}^{(n)}$ and the column of $\Psi_{0,0}^{(n)}$ have a single nonzero element each. In addition, the last row and the last column are all zero except for the diagonal element. Thus, expanding this $3(N+1)\times 3(N+1)$ determinant along those rows and columns, it is reduced to the following $3 N\times 3N$ determinant,
\begin{eqnarray}
&&\hspace{-1.0cm}\sigma_{n}=(-1)^{N+1} \times \nonumber \\
&&\hspace{-1.0cm}\left|
\begin{array}{cccc}
O_{N \times N} &
\begin{array}{cccc}
\Phi_{11}^{(n)} & \Phi_{12}^{(n)}  & \cdots & \Phi_{1, 2N}^{(n)}  \\
 \vdots &  \vdots  &   &  \vdots \\
\Phi_{2\widetilde{N}_{2}-1, 1}^{(n)} & \Phi_{2\widetilde{N}_{2}-1, 2}^{(n)}  & \cdots & \Phi_{2\widetilde{N}_{2}-1, 2N}^{(n)}  \\
\Phi_{21}^{(n)} & \Phi_{22}^{(n)}  & \cdots & \Phi_{2, 2N}^{(n)}  \\
 \vdots &  \vdots  &   &  \vdots \\
\Phi_{2\widetilde{N}_{1}, 1}^{(n)} & \Phi_{2\widetilde{N}_{1}, 2}^{(n)}  & \cdots & \Phi_{2\widetilde{N}_{1}, 2N}^{(n)}
\end{array} \\
 \begin{array}{cccccc}
\Psi_{11}^{(n)}  & \cdots & \Psi_{2\widetilde{M}_{2}-1, 1}^{(n)} & \Psi_{21}^{(n)}  & \cdots & \Psi_{2\widetilde{M}_{1}, 1}^{(n)} \\
\Psi_{12}^{(n)}  & \cdots & \Psi_{2\widetilde{M}_{2}-1, 2}^{(n)} & \Psi_{22}^{(n)}  & \cdots & \Psi_{2\widetilde{M}_{1}, 2}^{(n)}  \\
\vdots  &    & \vdots &   \vdots &  & \vdots  \\
\Psi_{1,2N}^{(n)}  & \cdots & \Psi_{2\widetilde{M}_{2}-1, 2N}^{(n)} & \Psi_{2, 2N}^{(n)}  & \cdots & \Psi_{2\widetilde{M}_{1}, 2N}^{(n)}
\end{array} & I_{2N \times 2N}
\end{array}
 \right|. \label{Second3Nby3N}
\end{eqnarray}

Now we use properties of Schur polynomials to transform the determinant (\ref{Second3Nby3N}) to (\ref{First3Nby3N}). For this purpose,
we consider Schur polynomials on vectors $\textbf{\emph{y}}=\textbf{\emph{x}}^{\pm}(n)+ \nu \textbf{\emph{s}}$ and $\hat{\textbf{\emph{y}}}=\textbf{\emph{x}}^{\pm}(n)+ (\nu+1) \textbf{\emph{s}}$, where $\textbf{\emph{x}}^{\pm}(n)$ and $\textbf{\emph{s}}$ are defined in equations (\ref{skrkexpcoeff})-(\ref{skrkexpcoeff2}). Since $\hat{\textbf{\emph{y}}}=\textbf{\emph{y}}+\textbf{\emph{s}}$, from the definition of Schur polynomials, we have
\begin{eqnarray}
&& \sum_{k=0}^{\infty}\left[S_{k}(\hat{\textbf{\emph{y}}})-S_{k}(\textbf{\emph{y}})\right]\lambda^k=
\exp\left(\sum_{j=1}^{\infty}\hat{y}_j\lambda^j\right)-\exp\left(\sum_{j=1}^{\infty}y_j\lambda^j\right)  \nonumber \\
&& =\exp\left(\sum_{j=1}^{\infty}y_j\lambda^j\right)\left[\exp\left(\sum_{j=1}^{\infty}s_j\lambda^j\right)-1\right]=
\sum_{k=0}^{\infty}S_{k}(\textbf{\emph{y}})\lambda^k \sum_{k=1}^{\infty}S_{k}(\textbf{\emph{s}})\lambda^k=
\sum_{k=1}^\infty\sum_{\begin{subarray}{l}
k_{1}+k_{2}=k\\
k_{1}\geq1, \hspace{0.07cm} k_{2}\geq 0
\end{subarray}}  S_{k_{1}}(\textbf{\emph{s}})S_{k_{2}}(\textbf{\emph{y}})\lambda^k.
\end{eqnarray}
From this, we get a relation between $S_{k}(\hat{\textbf{\emph{y}}})$ and $S_{k}(\textbf{\emph{y}})$ as
\[
S_{k}(\hat{\textbf{\emph{y}}})=S_{k}(\textbf{\emph{y}})+\sum_{\begin{subarray}{l}
k_{1}+k_{2}=k\\
k_{1}\geq1, \hspace{0.07cm} k_{2}\geq 0
\end{subarray}}  S_{k_{1}}(\textbf{\emph{s}})S_{k_{2}}(\textbf{\emph{y}}).
\]
In view that $S_{k_{1}}(\textbf{\emph{s}})$ are just constants, this means that polynomials $S_{k}(\hat{\textbf{\emph{y}}})$ can be expressed as linear combinations of $\{S_{j}(\textbf{\emph{y}}), 0\le j\le k\}$. In addition, since $s_1=s_3=\cdots=s_{odd}=0$,
\[
S_1(\textbf{\emph{s}})=S_3(\textbf{\emph{s}})=\cdots=S_{odd}(\textbf{\emph{s}})=0.
\]
Then, from the definition of functions $\Phi_{i,\nu}^{(n)}$ in Eq.~(\ref{phipsiexp}), we can directly show that
\[
\Phi_{i+1,\nu+1}^{(n)}=\frac{1}{2}\sum_{k=0}^{\left[\frac{i-\nu}{2}\right]}S_{2k}(\textbf{\emph{s}})\Phi_{i-2k,\nu}^{(n)}.
\]
Using this relation and performing simple row operations to the determinant (\ref{Second3Nby3N}), we can reduce its first $\widetilde{N}_{2}$ rows to the $(\widetilde{N}_{1}+1)$-th to $(\widetilde{N}_{1}+\widetilde{N}_{2})$-th rows of the determinant (\ref{First3Nby3N}), and reduce its $(\widetilde{N}_{2}+1)$-th to $(\widetilde{N}_{1}+\widetilde{N}_{2})$-th rows to the first $\widetilde{N}_{1}$ rows of the determinant (\ref{First3Nby3N}), plus a factor of $\frac{1}{2}$ on each element. Using similar treatments, we can reduce the first $\widetilde{M}_{2}$ columns of the determinant (\ref{Second3Nby3N}) to the $(\widetilde{M}_{1}+1)$-th to $(\widetilde{M}_{1}+\widetilde{M}_{2})$-th columns of the determinant (\ref{First3Nby3N}), and reduce the $(\widetilde{M}_{2}+1)$-th to $(\widetilde{M}_{1}+\widetilde{M}_{2})$-th columns of the determinant (\ref{Second3Nby3N}) to the first $\widetilde{M}_{1}$ columns of the determinant (\ref{First3Nby3N}), plus a factor of $\frac{1}{2}$ on each element. These factors of $\frac{1}{2}$ cancel out in the ratio of $\sigma_1/\sigma_0$, and thus solutions with $\left[ N_{1},  N_{2}, M_{1},  M_{2}\right]= \left[\widetilde{N}_{1}, \widetilde{N}_{2}, \widetilde{M}_{1},\widetilde{M}_{2}\right]$ and those with $\left[\widetilde{N}_{2}, \widetilde{N}_{1}+1, \widetilde{M}_{2}, \widetilde{M}_{1}+1\right]$ are equivalent to each other.

The second example of equivalency in Remark 5 is a direct consequence of the above equivalency and does not need proof.

Regarding the third example of equivalency in Remark 5, we notice that due to the above equivalency, solutions with indices $[N, K, N, K]$ and those with $[K-1, N, K-1, N]$ are equivalent. Also from the above equivalency, we see that solutions with indices $[K-1, N, K-1, N]$ and those with $[N-1, K-1, N-1, K-1]$ are equivalent. Combining these two results, we find that solutions with indices $[N, K, N, K]$ and those with $[N-1, K-1, N-1, K-1]$ are equivalent. Then, repeating this process, we can show that the third example of equivalency in Remark 5 is valid.


\begin{thebibliography}{10}

\bibitem{Akhmediev_2009}
Akhmediev N, Ankiewicz A and Taki M 2009 Waves that appear from nowhere and disappear
without a trace, Phys. Lett. A 373, 675.

\bibitem{Peregrine}
Peregrine D H 1983 Water waves, nonlinear Schrodinger equations and their solutions, J. Aust.
Math. Soc. B 25, 16-43.

\bibitem{Ocean_rogue_review}
Dysthe K, Krogstad H.E. and M\"uller P 2008, Oceanic Rogue Waves,
Annu. Rev. Fluid Mech. 40, 287-310.

\bibitem{Pelinovsky_book}
Kharif C, Pelinovsky E and Slunyaev A 2009 Rogue Waves in the Ocean (Springer, Berlin).

\bibitem{Solli_Nature}
Solli D R, Ropers C, Koonath P and Jalali B 2007 Optical rogue waves, Nature 450, 1054-1057.

\bibitem{Wabnitz_book}
Wabnitz S. (Ed.) 2017 \emph{Nonlinear Guided Wave Optics: A testbed for extreme waves} (IOP Publishing, Bristol, UK).

\bibitem{Tank1}
Chabchoub A., Hoffmann N.P. and Akhmediev N. 2011 Rogue wave observation in a water wave tank, Phys.
Rev. Lett. 106, 204502.

\bibitem{Tank2}
Chabchoub, A., Hoffmann, N., Onorato, M., Slunyaev, A., Sergeeva, A., Pelinovsky, E., Akhmediev,
N. 2012 Observation of a hierarchy of up to fifth-order rogue waves in a water tank. Phys. Rev. E 86, 056601.

\bibitem{Fiber1}
Kibler, B., Fatome, J., Finot, C., Millot, G., Dias, F., Genty, G., Akhmediev, N., Dudley, J.M. 2010 The
Peregrine soliton in nonlinear fibre optics. Nat. Phys. 6, 790-795.

\bibitem{Fiber2}
Frisquet, B., Kibler, B., Morin, P., Baronio, F., Conforti, M., Millot, G., Wabnitz, S. 2016 Optical dark rogue
wave. Sci. Rep. 6, 20785.

\bibitem{Fiber3} Baronio, F., Frisquet, B., Chen, S., Millot, G., Wabnitz, S., Kibler, B. 2018 Observation of a group of dark
rogue waves in a telecommunication optical fiber. Phys. Rev. A 97, 013852.

\bibitem{AAS2009}
Akhmediev N, Ankiewicz A and Soto-Crespo J.M. 2009 Rogue waves and rational solutions of the nonlinear Schr\"odinger equation, Phys. Rev. E 80, 026601.

\bibitem{DGKM2010}
Dubard P, Gaillard P, Klein C. and Matveev V.B. 2010 On multi-rogue wave solutions of the NLS
equation and positon solutions of the KdV equation, Eur. Phys. J. Spec. Top.  185, 247-258.

\bibitem{ACA2010}
Ankiewicz A, Clarkson P.A. and Akhmediev N 2010 Rogue waves, rational solutions, the patterns of their zeros and integral relations, J. Phys. A 43, 122002.

\bibitem{DPMB2011}
Dubard P and Matveev V.B. 2011 Multi-rogue waves solutions to the focusing NLS equation and the KP-I equation,
Nat. Hazards Earth Syst. Sci. 11, 667-672.

\bibitem{KAAN2011}
Kedziora D.J, Ankiewicz A and Akhmediev N, 2011 Circular rogue wave clusters, Phys. Rev. E  84, 056611.

\bibitem{GLML2012}
Guo B.L, Ling L.M and Liu Q.P 2012 Nonlinear Schrodinger equation: generalized Darboux transformation and rogue wave solutions, Phys. Rev. E 85, 026607.

\bibitem{OhtaJY2012}
Ohta Y and Yang J 2012 General high-order rogue waves and their dynamics in the nonlinear
Schrodinger equation, Proc. R. Soc. Lond. A 468, 1716-1740.

\bibitem{DPMVB2013}
Dubard P and Matveev V.B 2013 Multi-rogue waves solutions: from the NLS to the KP-I equation, Nonlinearity 26,  R93-R125.

\bibitem{XuHW2011}
Xu S, He J.S and Wang L 2011 The Darboux transformation of the derivative nonlinear Schr\"odinger
equation, J. Phys. A 44, 305203.

\bibitem{GLML2013}
Guo B.L, Ling L.M and Liu Q.P 2013 High-order solutions and generalized Darboux transformations
of derivative nonlinear Schr\"odinger equations, Stud. Appl. Math. 130,  317-344.

\bibitem{BCDL2013}
Baronio F, Conforti M, Degasperis A and Lombardo S 2013 Rogue waves emerging from the
resonant interaction of three waves, Phys. Rev. Lett. 111, 114101.

\bibitem{OhtaJKY2012}
Ohta Y and Yang J 2012 Rogue waves in the Davey-Stewartson I equation, Phys. Rev. E 86, 036604.

\bibitem{OhtaJKY2013}
Ohta Y and Yang J 2013 Dynamics of rogue waves in the Davey-Stewartson II equation, J. Phys. A 46, 105202.

\bibitem{AANJM2010}
Ankiewicz A, Akhmediev N and Soto-Crespo J.M 2010 Discrete rogue waves of the Ablowitz-Ladik and Hirota equations, Phys. Rev. E 82,  026602.

\bibitem{OhtaJKY2014}
Ohta Y and Yang J 2014 General rogue waves in the focusing and defocusing Ablowitz-Ladik equations, J. Phys. A 47, 255201.

\bibitem{ASAN2010}
Ankiewicz A, Soto-Crespo J.M and Akhmediev N 2010 Rogue waves and rational solutions of the Hirota equation, Phys. Rev. E 81, 046602.

\bibitem{TaoHe2012}
Tao Y.S and He J.S 2012 Multisolitons, breathers, and rogue waves for the Hirota equation generated
by the Darboux transformation, Phys. Rev. E 85, 026601.

\bibitem{BDCW2012}
Baronio F, Degasperis A, Conforti M and Wabnitz S 2012 Solutions of the vector
nonlinear Schr\"odinger equations: evidence for deterministic rogue waves, Phys. Rev. Lett.
109, 044102.

\bibitem{ManakovDark}
Baronio F, Conforti M, Degasperis A, Lombardo S, Onorato M and Wabnitz S 2014
Vector Rogue Waves and Baseband Modulation Instability in the Defocusing Regime, Phys. Rev. Lett. 113, 034101.

\bibitem{Chow}
Chow K.W, Chan H.N, Kedziora D.J and Grimshaw R.H.J 2013 Rogue wave modes for the long wave�short wave resonance model, J. Phys.
Soc. Jpn. 82, 074001.

\bibitem{PSLM2013}
Priya N.V, Senthilvelan M and Lakshmanan M 2013 Akhmediev breathers, Ma solitons, and general
breathers from rogue waves: a case study in the Manakov system, Phys. Rev. E 88, 022918.

\bibitem{Grimshaw_rogue}
Mu G, Qin Z and Grimshaw R 2015 Dynamics of rogue waves on a multi-soliton background in a vector nonlinear Schr\"odinger equation, SIAM J. Appl. Math. 75, 1-20.

\bibitem{MuQin2016}
Mu G and Qin Z 2016 Dynamic patterns of high-order rogue waves for Sasa-Satsuma equation, Nonlinear Anal. Real World Appl. 31, 179-209.

\bibitem{LLMSasa2016}
Ling L.M 2016 The algebraic representation for high order solution of Sasa-Satsuma equation, Discrete Continuous Dyn Syst Ser B 9, 1975-2010.

\bibitem{LLMFZ2016}
Ling L.M, Feng B.F and Zhu Z 2016 Multi-soliton, multi-breather and higher order rogue wave solutions to the complex short pulse equation,  Physica D 327, 13-29.

\bibitem{Degasperis1}
Degasperis A and Lombardo S 2013 Rational solitons of wave resonant-interaction models, Phys. Rev. E 88, 052914.

\bibitem{Degasperis2}
Degasperis A. and Lombardo S 2016 Integrability in Action: Solitons, Instability and Rogue Waves, In: M. Onorato, S. Resitori and F. Baronio (eds) \emph{Rogue and Shock Waves in Nonlinear Dispersive Media}, Lecture Notes in Physics, vol 926, pp. 23-53, Springer, Cham.

\bibitem{Ablowitz_PRL}
Ablowitz M.J and Musslimani Z.H 2013 Integrable nonlocal nonlinear Schr\"odinger equation, Phys. Rev. Lett. 110, 064105.

\bibitem{Yang_PTreview}
Konotop V.V, Yang J and Zezyulin D.A 2016 Nonlinear waves in PT-symmetric systems. Rev. Mod. Phys.
88, 035002.

\bibitem{magnetics} Gadzhimuradov T.A and Agalarov A.M 2016 Towards a gauge-equivalent magnetic
structure of the nonlocal nonlinear Schr\"odinger equation, Phys. Rev. A 93, 062124.

\bibitem{Ablowitz_nonlinearity}
Ablowitz M.J and Musslimani Z.H 2016 Inverse scattering transform for the integrable nonlocal nonlinear
Schrodinger equation. Nonlinearity 29, 915-946.

\bibitem{WYY2016}
Wen X.Y, Yan Z and Yang Y 2016 Dynamics of higher-order rational solitons for the nonlocal nonlinear Schrodinger equation with the self-induced parity-time-symmetric potential, Chaos 26, 063123.

\bibitem{HXLM2016}
Huang X and Ling L.M 2016 Soliton solutions for the nonlocal nonlinear Schrodinger equation, Eur. Phys. J. Plus 131, 148.

\bibitem{Gerdjikov2017} Gerdjikov V.S and Saxena A 2017 Complete integrability of nonlocal nonlinear Schr\"odinger equation,
J. Math. Phys. 58, 013502.

\bibitem{YangPTNLSsoliton} Yang J 2018 General N-solitons and their dynamics in several nonlocal nonlinear Schroedinger equations, Phys. Lett. A 383, 328-337.

\bibitem{FengPTNLSsoliton}
Feng B.F, Luo X.D, Ablowitz M.J and Musslimani Z.H 2018 General soliton solution to a nonlocal nonlinear
Schr\"odinger equation with zero and nonzero boundary conditions, Nonlinearity 31, 5385.

\bibitem{YangYang2018}
Yang  B and Yang J 2019 Rogue waves in the nonlocal \PT-symmetric nonlinear
Schr\"{o}dinger equation, Lett. Math. Phys. 109, 945-973. 

\bibitem{Hirota}
Hirota R 2004 The direct method in soliton theory (Cambridge University Press, Cambridge).

\bibitem{Chen_Juntao}
Chen J, Chen Y, Feng B.F, Maruno K and Ohta Y 2018 General high-order rogue waves of the (1+1)-dimensional Yajima-Oikawa system, J. Phys. Soc. Jpn. 87, 094007.

\bibitem{XiaoeYong2018}
Zhang X and Chen Y 2018 General high-order rogue waves to nonlinear Schr\"{o}dinger-Boussinesq equation with the dynamical analysis, Nonlinear Dynamics 93, 2169-2184.

\bibitem{SunLian2018}
Sun B and Lian Z 2018 Rogue waves in the multicomponent Mel'nikov system and multicomponent Schr\"{o}dinger-Boussinesq system, Pramana 90, 23.

\end{thebibliography}
\end{document}